\documentclass[aps,prb,superscriptaddress,showpacs,twocolumn]{revtex4}
\usepackage{graphics,graphicx}
\usepackage{amssymb}
\usepackage{hyperref}
\def\vm{{\bf m}}
\def\vn{{\bf n}}
\def\vp{{\bf p}}
\def\vs{{\bf s}}
\def\vv{{\bf v}}
\def\vx{{\bf x}}
\def\vy{{\bf y}}
\def\vz{{\bf z}}
\newcommand{\sgn}{\mbox{sgn}}

\renewcommand{\Re}{\mathfrak{Re}\,}
\def\ket#1{\mbox{$\displaystyle\vert\,#1\,\rangle$}}
\def\bra#1{\mbox{$\displaystyle\langle\,#1\,\vert$}}

\newcommand{\be}{\begin{equation}}
\newcommand{\ee}{\end{equation}}
\newcommand{\ber}{\begin{eqnarray}}
\newcommand{\eer}{\end{eqnarray}}
\def\onlinecite#1{\cite{#1}}

\begin{document}
\title{Theory of Nonequilibrium Spin Transport and Spin Transfer Torque
       in Superconducting-Ferromagnetic Nanostructures}
\author{Erhai Zhao}
\affiliation{Department of Physics and Astronomy, University of Pittsburgh, Pittsburgh, Pennsylvania 15260, USA}
\author{J. A. Sauls}
\affiliation{Department of Physics and Astronomy, Northwestern University, Evanston, Illinois 60208, USA}
\pacs{72.25.Mk, 74.50.+r}
\date{\today}
\begin{abstract}
Spin transport currents and the spin-transfer torques in
voltage-biased superconducting-ferromagnetic nanopillars (SFNFS
point contacts) are computed. We develop and implement an algorithm based on
the Ricatti formulation of the quasiclassical theory of superconductivity
to solve the time-dependent boundary conditions for the
nonequilibrium Green's functions for spin transport through the
ferromagnetic interfaces. A
signature of the nonequilibrium torque is a component perpendicular
to the plane spanned by the two ferromagnetic moments. The
perpendicular component is absent in normal-metal-ferromagnetic
nanopillars (NFNFN) contacts, but is shown to have the same order of
magnitude as the in-plane torque for non-equilibrium SFNFS contacts.
The out-of-plane torque is due to the rotation of quasiparticle spin
by the exchange fields of the ferromagnetic layers. In the ballistic
limit the equilibrium torque is related to the spectrum of
spin-polarized Andreev bound states, while the {\sl ac} component,
for small bias voltages, is determined by the nearly adiabatic dynamics of
the Andreev bound states. The nonlinear voltage dependence of the
non-equilibrium torque, including the subharmonic gap structure and
the high-voltage asymptotics, is attributed to the interplay between
multiple Andreev reflections, spin filtering and spin mixing. These
properties of spin angular momentum transport may be exploited to
control the state of nanomagnets.
\end{abstract}
\maketitle
\section{Introduction}

Spin-polarized current passing through a ferromagnet (F) can
transfer spin angular momentum to the ferromagnet and exert a torque
on the magnetic moment \cite{slon,berger,kelly_review,stiles_review}. 
This provides a mechanism
for manipulating the state of nanomagnets. For example,
magnetization precession and reversal driven by the spin-transfer
torque has been observed in ferromagnet-normal metal
multilayers \cite{tsoi,myers}. A promising multilayer geometry for
device applications, such as random access memory and microwave
oscillators, is the so-called ``magnetic nanopillar'' consisting of
a ferromagnet-normal-metal-ferromagnet (FNF) trilayer connected to
normal metal (N) electrodes \cite{nature03,science05}. The typical
thickness of each layer is several nanometers, and the diameter of
the pillar is of the order of 50-100 nm.

Slonczewski proposed a unified description of equilibrium and
nonequilibrium spin-transfer torque in magnetic nanopillars based on
the dynamics of the spin current \cite{slon89}. The spin current is
not conserved across the ferromagnetic interface, so the
spin-transfer torque on each ferromagnetic (F) layer equals the
net spin flux it absorbs. In thermal equilibrium a persistent spin
current flows in the N layer between the two ferromagnets. The spin
current is related to the formation of spin-polarized states due to
the spin-dependent confinement potential of the FNF
trilayer \cite{ortega,stiles93,bruno95}. The equilibrium current is
responsible for the exchange torque between the two F layers. This
torque prefers either parallel or anti-parallel alignment of the
magnetic moments, depending on the thickness of the N
layer \cite{gru,parkin}. Voltage-biasing an FNF trilayer results in a
nonequilibrium spin current in the N layer, as well as in the
electrodes. The resulting torque on each F layer is proportional to
the voltage.

A scattering theory for computing spin current and spin-transfer
torque in normal metal-ferromagnet hybrid structures has been
developed by several authors \cite{circuit,waintal00,hei2008,stiles02}.
Within this approach the ferromagnetic layers are treated as sharp
interfaces between normal conducting leads. Scattering matrices
describing the transmission and reflection of electronic
excitations, together with their spectra and distribution functions
in the electrodes, determine the charge and spin transport.

Waintal and Brouwer pointed out that spin angular momentum transfer
acquires new features in magnetic nanopillars connected to
superconducting (S) electrodes \cite{waintal01,waintal02}. For
example, they showed that the nonequilibrium torque in NFNFS
junctions may favor a perpendicular configuration of the two moments
due to Andreev reflection at the NS interface for voltages below the
superconducting gap, $eV < \Delta$ \cite{waintal01}. They also showed
that the equilibrium torque in SFNFS junctions depends on the condensate
phase difference, $\phi$, between the superconducting
leads \cite{waintal02}. These features are consequences of
energy-dependent, phase-sensitive scattering of the electronic
excitations at the ferromagnetic interfaces. Fabrication and control
of superconducting-ferromagnetic nanopillars appears to be within
the capability of current technology \cite{bell}. For example, the
{\sl dc} Josephson effect in SFNFS junctions has been reported in
Ref. \onlinecite{bell}.

In order to better understand, design and develop superconducting
spin-transfer devices, a quantitative theory of spin transport and
spin angular momentum transfer under nonequilibrium conditions is
needed. In this paper we develop the theory of nonequilibrium spin
transport and spin-transfer torque in voltage-biased SFNFS point
contacts, i.e. superconducting nanopillars with diameter smaller
than the superconducting coherence length of the electrodes. Under
voltage bias, $V$, the phase difference, $\phi$, increases with time
at a rate given by the Josephson frequency, $\omega_J =2eV/\hbar$.
In response, the spin current and spin-transfer torque oscillate
with time. In general, the time-dependent torque excites the
ferromagnet by setting the magnetic moments into motion, which in
turn affects the spin transport. A full theory of the coupled
dynamics of the ferromagnetic moments and the spin current is
outside the scope of this work. In this paper we assume the
ferromagnetic moments do not change direction on the time scale set
by the Josephson frequency, $T_J=2\pi/\omega_J$. This assumption is
justified for point contacts over a wide range of voltages because
the characteristic frequency for the magnetization dynamics is set
by the magnitude of the torque which scales linearly with the
cross-sectional area of the contact, $\mathcal{A}$. For junctions
with sufficiently large cross section, our assumption holds only for
high voltage bias. The quasi-static approximation eventually breaks
down when the bias voltage is sufficiently small compared to the
superconducting gap that the magnetization dynamics is comparable
to, or faster, than the phase dynamics. Some aspects of this limit
are discussed in Ref. \onlinecite{waintal02}.

Our theoretical approach is conceptually similar to the scattering
theory for spin-transfer torque in normal-metal-ferromagnet
hybrids \cite{circuit,waintal00}. We develop a scattering formulation
for spin transport which includes the effects of particle-hole
coherence arising from the superconducting leads. Our method is
based on the Ricatti formulation of the quasiclassical theory of
superconductivity \cite{schopol95,schopol98,eschrig99,shelankov00,eschrig00}
and is explained in Sec. \ref{sec_methods}. We introduce the
particle-hole coherence functions, the quasiparticle distribution
functions and the boundary conditions obeyed by these functions at
ferromagnetic interfaces. We also discuss the scattering matrices
that describe the ferromagnetic interfaces; these matrices
incorporate the key effects of spin-dependent scattering: {\sl spin
mixing and spin filtering}. In this section we also summarize the
method we use to solve the interface boundary conditions. In Sec.
\ref{sec_equilibrium-torque}, we discuss the theory of the
equilibrium torque in SFNFS contacts. We identify the microscopic
scattering processes that give rise to the out-of-plane torque and
derive an analytical result for the spin-transfer torque in
terms of the spectrum of Andreev bound states at the point contact.
In Sec. \ref{sec_nonequilibrium-torque}, we report results for the
voltage, temperature, and misalignment angle dependences of the {\sl
dc} and {\sl ac} components of the spin current and spin-transfer
torque in voltage-biased SFNFS nanopillars. We summarize our results in Sec.
\ref{sec_discussion}. Mathematical
details of the solution of the time-dependent boundary conditions
are presented in the appendix.

\section{Theory}\label{sec_methods}

A powerful formalism to study time-dependent nonequilibrium
transport in superconducting heterostructures is provided by the
quasiclassical theory of superconductivity \cite{serene83,rammer86}.
This theory describes phenomena on length scales larger than the
Fermi wavelength, $\lambda_f$, and on time scales long compared to
the inverse Fermi energy, $\hbar/E_f$. The central equations of the
theory are a set of transport equations which govern the spatial and
temporal variations of quasiclassical Green's functions. Strong
localized potentials such as surfaces and interfaces are
characterized by a scattering matrix, an interface ``S-matrix'',
which defines the probability amplitudes for scattering between
incident and outgoing normal-state particle and hole quasiparticles.
The interface S-matrix is the key input to the boundary conditions
for the quasiclassical Green's
functions \cite{zaitsev84,kieselmann85,millis88}.

Although these boundary conditions have been used to investigate the
proximity effect and current-voltage characteristics of tunnel
junctions and point contacts, their application has been somewhat
limited. The boundary conditions are nonlinear, possess spurious
unphysical solutions which have to be discarded and are
time-consuming to implement numerically under non-equilibrium
conditions. An improved formulation of the quasiclassical boundary
conditions was achieved by Eschrig\cite{eschrig00} for non-magnetic
interfaces by reducing the boundary conditions for the
quasiclassical Nambu-matrix propagators to boundary conditions for
the particle-hole coherence functions, $\gamma^{R/A}$ and
$\tilde{\gamma}^{R/A}$, and the distribution functions for the
particle-like, $x^K$, and hole-like, $\tilde{x}^K$, excitations (we
follow the notation of Ref. \onlinecite{eschrig00}). These
components are $2\times 2$ spin matrices defined on classical
trajectories labelled by the Fermi velocity, $\mathbf{v}_f$; and are
functions of spatial position, $\mathbf{R}$, Fermi momentum
$\mathbf{p}_f$, excitation energy $\epsilon$, and time, $t$.
Note that pairs of particle-like and hole-like components, e.g.
$\tilde{q}$ and $q$, are related by the conjugation symmetry,
$\tilde{q}(\mathbf{p}_f,\epsilon)={q}^*(-\mathbf{p}_f,-\epsilon)$.

The equilibrium coherence functions determine the relative amplitude
of normal-state particle and hole states that define the Bogoliubov
exictations for the superconducting state. The pair of coherence
functions also combine to determine local quasiparticle excitation
spectrum, i.e.
$N(\mathbf{p}_f,\epsilon)= \Re[(1+\gamma^R\tilde{\gamma}^R)/
    (1-\gamma^R\tilde{\gamma}^R)]$.
For non-equilibrium conditions, and in the clean limit, the
coherence functions also determine the probability amplitudes for
branch conversion from electron-like to hole-like quasiparticles,
and vice versa. The distribution functions determine the occupation
probabilities for the particle- and hole-like excitations. We refer
to both the coherence functions and distribution functions as
Riccati amplitudes, since all of these functions obey
Riccati-type \cite{ried} differential
equations \cite{schopol95,eschrig99,shelankov00,eschrig00}. These
equations are supplemented by boundary conditions at an interface,
and by asymptotic conditions deep in the bulk electrodes. For our
purposes a complete set of boundary conditions for the
non-equilibrium Riccati amplitudes were recently derived for
spin-active interfaces \cite{fogelstrom,zls}. In contrast to the
nonlinear boundary conditions for quasiclassical
propagators \cite{millis88} the boundary conditions for Riccati
amplitudes are easier to solve and free of unphysical solutions.
Thus, the Riccati formulation of quasiclassical theory provides an
efficient method for finding the nonequilibrium Green's functions
near spin-active interfaces \cite{zls}.

Several authors \cite{ashida,shelankov00,brinkman00,gala,ozana02}
have noted that caution must be exercised when applying the
quasiclassical transport equations and related boundary conditions
to multi-barrier proximity structures even for structures with
dimensions that are large compared to the Fermi wavelength. When
there are multiple interfaces present the quasiclassical formulation
can lead to errors that exceed the small expansion
parameters that are the basis for the quasiclassical approximation,
i.e $1/k_f\xi_0 \ll 1$, etc.
This breakdown of quasiclassical approximation is believed due to the
formation of closed trajectories that give rise to constructive,
quantum interference terms resulting from scattering (for
normal-state quasiparticles) by several different interfaces which
invalidates the normalization condition for quasiclassical Green's
functions \cite{shelankov00,ozana02}.
However, for layered structures with two interfaces the
quasiclassical theory yields correct results for junctions with many
transmission channels. The quantum interference effect,
characterized by channel-dependent phase factors
averages out when summed over many conduction
channels. This has been verified by comparing the results of
quasiclassical theory and microscopic Bogoliubov-de Gennes or Gorkov
theory on the dc Josephson effect in double barrier SINIS junctions
(I stands for an insulator layer) \cite{brinkman00,gala,ozana02}. In
Ref. \onlinecite{brinkman03}, quasiclassical theory was applied to
study nonequilibrium charge transport in diffusive double-barrier
SINIS junctions by solving the boundary conditions at the two
insulator interfaces. Our method of computing spin currents and the spin-transfer
torque in magnetic nanopillars is analogous to that developed in
Ref. \onlinecite{brinkman03}. Given that magnetic nanopillars have
many conduction channels \cite{circuit,waintal00}, the
quasiclassical approach is justified.

\begin{figure}
  \includegraphics[width=3.5in]{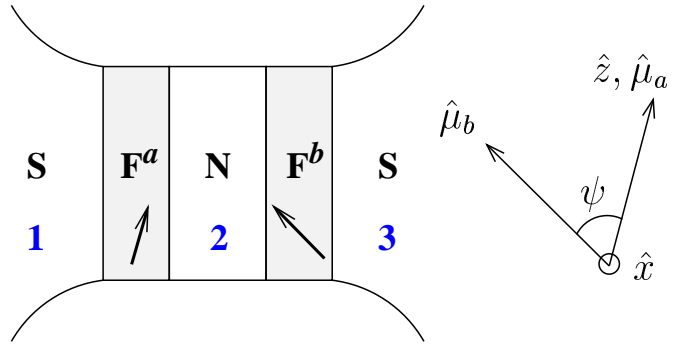}
  \caption{Schematic of superconducting magnetic nanopillar. The left electrode, middle
  spacer, and right electrode are also called region 1, 2, and 3,
  respectively. The two ferromagnetic layers are labeled by indices
  $a$ and $b$. Their magnetization directions, $\hat{\mu}_a$ and $\hat{\mu}_b$,
  are in general mis-aligned; $\hat{\mu}_a$ is along the $z$ axis, while $\hat{\mu}_b$ is at
 polar angle $\psi$ in the $yz$ plane. The $x$ axis is along $\hat{\mu}_a\times
 \hat{\mu}_{b}$, and the diameter of the pillar is $d\ll \xi_0$.}
  \label{junction}
\end{figure}

In this article we apply the Riccati method to study time-dependent
transport in superconducting-ferromagnetic nanopillars (SFNFS
junctions). The junction geometry and coordinate system are shown in
Fig. \ref{junction}. Two single-domain, nano-scale ferromagnets,
F$^a$ and F$^b$, are separated by a normal-metal spacer, N. The FNF
trilayer is connected in a circuit with electrodes (S) made of
conventional $s$-wave superconductors. The ferromagnetic interfaces,
F$^a$ and F$^b$, are characterized by their scattering matrices,
$\mathbb{S}^a$ and $\mathbb{S}^b$, respectively, that couple
normal-state excitations in adjacent electrodes.

For simplicity we are assume the ferromagnets to be identical except
that their magnetic moments are mis-aligned by an angle $\psi$. The
thickness of each layer, as well as the diameter of each
nano-pillar, is assumed to be much smaller than the superconducting
coherence length, $\xi_0=\hbar v_f/2\pi k_B T_c$, and the inelastic
mean free path, $\ell_{in}$. These criteria are easily met, for
example, with Al electrodes. The N layer thickness $L$ is assumed to
be larger than the Fermi wavelength but still thin compared to the
superconducting coherence length, $\lambda_f\ll L\ll\xi_0$. We also
assume the leads and the trilayer are in the clean limit and the
scattering at F$^a$ and F$^b$ is specular. We choose the
magnetization direction, $\hat{\mu}_a$, as the quantization axis
($z$ axis) for spin. The magnetic moments, $\hat{\mu}_a$ and
$\hat{\mu}_b$, span the $yz$ plane, while the $x$ axis is
perpendicular to both $\hat{\mu}_a$ and $\hat{\mu}_b$. As discussed
above, we assume $\hat{\mu}_a$ and $\hat{\mu}_b$ are static on the
Josephson time scale, $T_J$.

\begin{figure}
\includegraphics[width=3.5in]{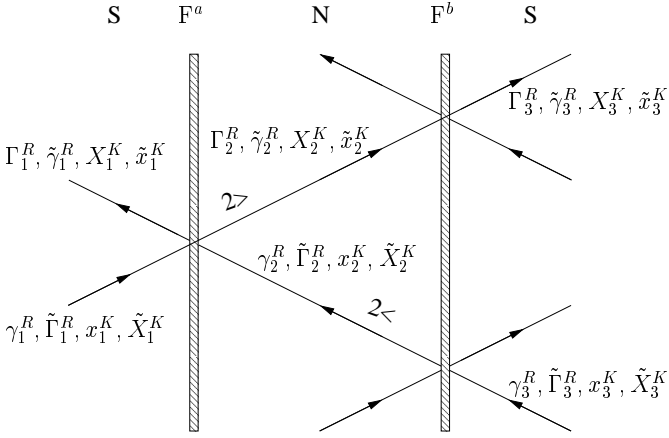}
\caption{Coupled scattering trajectories (directed lines) and the
corresponding Riccati amplitudes for an SFNFS structure in the clean
limit. The arrows indicate the direction of the Fermi momentum for
quasiparticles defined on each trajectory. The right (left) going
trajectories in the N region are labeled by 2${>}$ (2${<}$).}
\label{tri_bc}
\end{figure}

Figure \ref{tri_bc} shows the Riccati amplitudes defined on a set of
scattering trajectories near the ferromagnetic interfaces. The
Riccati amplitudes in the superconducting electrodes are classified
into ``incoming" Riccati amplitudes denoted by lower case symbols,
$\{\gamma^{R,A}_j,\tilde{\gamma}^{R,A}_j,x^K_j,\tilde{x}^K_j\}$, and
``outgoing" Riccati amplitudes denoted by upper case symbols,
$\{\Gamma^{R,A}_j,\tilde{\Gamma}^{R,A}_j,X^K_j,\tilde{X}^K_j\}$,
$j=1,3$ \cite{eschrig00,zls}. The amplitudes in the N region are also
labeled in this way with ``incoming" and ``outgoing" groups defined
with respect to the interface with F$^a$.

Proximity effects, i.e. the appearance of superconducting
correlations in the N layer, as well as the suppression of
superconductivity in the electrodes by the FNF trilayer, are encoded
in the local particle-hole coherence functions. Charge and spin
currents in each region, $i=1,2,3$, are determined by the local
distribution functions together with the coherence functions. For
SFNFS junctions under a constant voltage bias the phase difference
across the junction evolves according to the Josephson-Anderson
relation, $\dot{\phi}(t)=\omega_J=2eV/\hbar$. Thus, all local
Riccati amplitudes defined above are time-dependent. However, we can
neglect spatial variations of the Riccati amplitudes within the N
region. For example, the coherence function, $\Gamma^R_2$, in the N
layer varies in space as
$\Gamma^R_2(z')\sim\Gamma^R_2(0)e^{ik_Tz'}$, where
$k_T=2\epsilon/v_f\ll k_f$ is the Tomasch wave vector and $z'$ is
the distance from F$^a$ along trajectory 2$>$. Since $z'$ is
typically of order the N layer thickness (several nanometers) the
phase accumulation $k_Tz'$ is negligible, except for grazing
trajectories that are irrelevant for most transport phenomena across
the trilayer. Similarly the distribution function, $X^K_2$, varies
as $X^K_2(z')\sim X^K_2(0)e^{i\omega z'/v_f}$, where $\omega$ is of
the order of the Josephson frequency $\omega_J\ll E_f$, and barely
changes except for grazing trajectories. Thus, for thin N layers
determining the local Riccati amplitudes is reduced to a
simultaneous solution to the boundary conditions at F$^a$ and
F$^b$.

\subsection*{Boundary conditions}

Boundary conditions at F$^a$ and F$^b$ connect the ``outgoing"
Ricatti amplitudes to the ``incoming" Ricatti amplitudes via the
interface scattering matrices \cite{zls}. For voltage-biased
junctions these boundary conditions describe the spin-dependent
inelastic reflection and transmission of quasiparticles, loosely
referred to as ``multiple Andreev reflection'' (MAR).
The summation of the multiple scattering processes are contained in
the boundary conditions, which for the retarded Ricatti amplitudes
are
\begin{eqnarray}
\Gamma^R_1 &= r^a_{1}\circ\gamma^R_1
(\underline{S}^a_{11})^{\dagger} + t^a_{1}\circ\gamma^R_2
(\underline{S}^a_{12})^{\dagger}\,,
\label{GM1}\\
\Gamma^R_2&=r^a_2\circ \gamma^R_2 (\underline{S}^a_{22})^{\dagger}
+t^a_2\circ \gamma^R_1   (\underline{S}^a_{21})^{\dagger}\,,
\label{GM2}\\
\gamma^R_2&=r^b_2\circ \Gamma^R_2 (\underline{S}^b_{22})^{\dagger}
+t^b_2\circ \gamma^R_3   (\underline{S}^b_{23})^{\dagger}\,,
\label{gm2}\\
\Gamma^R_3 &= r^b_{3}\circ\gamma^R_3
(\underline{S}^b_{33})^{\dagger} + t^b_{3}\circ\Gamma^R_2
(\underline{S}^{b}_{32})^{\dagger}\,. \label{GM3}
\end{eqnarray}
Matrix multiplication in spin space is implied, and the $\circ$
operator denotes the folding product, which in the time domain is
defined as,
\begin{equation}
[A\circ B](t_1,t_2)=\int_{-\infty}^{+\infty}dt\;A(t_1,t)B(t,t_2).
\end{equation}

\begin{table}
\begin{tabular}{ccc}
Amplitude            & Event                & Diagram \\
\hline \hline
$\Gamma^R_1$         & $1h>\rightarrow 1e<$ & \includegraphics[width=0.45cm]{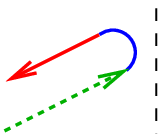} \\
$\gamma^R_1$         & $1h<\rightarrow 1e>$ & \includegraphics[width=0.40cm]{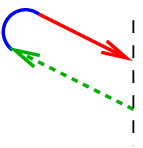} \\
$\gamma^R_2$         & $2h>\rightarrow 2e<$ & \includegraphics[width=0.40cm]{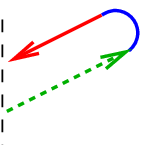} \\
$\tilde{\gamma}^R_2$ & $2e>\rightarrow 2h<$ & \includegraphics[width=0.40cm]{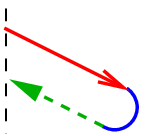} \\
$r^a_{1}$            & $1e>\rightarrow 1e<$ & \includegraphics[width=0.40cm]{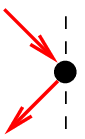} \\
$t^a_{1}$            & $2e<\rightarrow 1e<$ & \includegraphics[width=0.45cm]{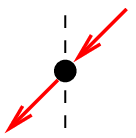} \\
\hline
\end{tabular}
\caption[]{Amplitudes for reflection and transmission of electrons
and holes for the processes contributing to Eq. \ref{GM1}. $h$ ($e$)
stands for hole (electron), and $>$ ($<$) represents right (left)
moving. For example, $1h> \rightarrow 1e<$ means a right moving hole
in region 1 converts into a left moving electron in region 1. }
\label{table_amplitudes}
\end{table}

\begin{figure}[h]
\includegraphics[width=2.5in]{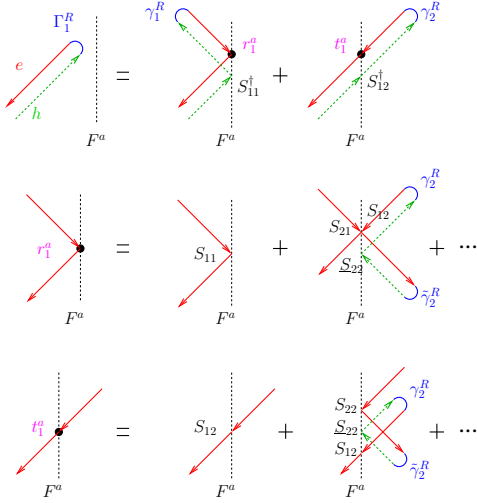}
\caption{Diagrams for interface scattering with branch conversion
from hole to particle. The full scattering amplitude includes
Andreev reflection and transmission with branch conversion to all
orders in the normal state transmission ($S_{12}, S_{21}$) and
reflection amplitudes ($S_{11}, S_{22})$.}
\label{diagrams_interface}
\end{figure}

These boundary conditions can be expressed in terms of a set of
rules for computing the branch-conversion scattering amplitudes near
the interface. Table \ref{table_amplitudes} lists the scattering
amplitudes that determine the outgoing Ricatti amplitude,
$\Gamma^R_1$, for electrode $1$. This is the amplitude for
retro-reflection by the interface with branch conversion from a hole-like
to a particle-like excitation. The diagrammatic representation of
the lowest order elementary scattering processes that contribute to
$\Gamma^R_1$ are shown in Fig. \ref{diagrams_interface}.

The transmission ($t$) and reflection ($r$) amplitudes in Eqs.
\ref{GM1}-\ref{GM3} are calculated from the normal-state,
branch-conserving scattering amplitudes
($S^{a/b}_{ij},\underline{S}^{a/b}_{ij}$) and elementary
branch-conversion amplitudes ($\gamma^R_j,\tilde{\gamma}^R_j$) by
summing the amplitudes for all multiple scattering processes. In
particular, for transmission and reflection from the central
electrode ($2$) through F$^{a/b}$,
\begin{eqnarray}
t^a_{2} &=&
\left[(S^a_{21})^{\dagger}-\beta^a_{11}\circ(\beta^{a}_{12})^{-1}
(S^a_{22})^{\dagger} \right]^{-1}\,,
\label{ta2}\\
r^a_2 &=& -t^a_{2}\circ\beta^a_{11}\circ(\beta^{a}_{12})^{-1}\,,
\label{ra2}\\
t^b_{2} &=&
\left[(S^b_{23})^{\dagger}-\beta^b_{33}\circ(\beta^{b}_{32})^{-1}
(S^b_{22})^{\dagger} \right]^{-1}\,,
\label{tb2}\\
r^b_2 &=& -t^b_{2}\circ\beta^b_{33}\circ(\beta^{b}_{32})^{-1} \,,
\label{rb2}
\end{eqnarray}
where the coefficients $\beta^a_{ij}$ are defined by
\begin{eqnarray}
\beta^a_{ij} &=& (S^a_{ij})^{\dagger}-\gamma^R_j
(\underline{S}^a_{ij})^{\dagger}\circ \tilde{\gamma}^R_i\,,
\label{beta} \\
\tilde{\beta}^a_{ij} &=& \underline{S}^a_{ji}-\tilde{\gamma}^R_j
\underline{S}^a_{ji}\circ \gamma^R_i,\; i,j=1,2 \,. \label{betat}
\end{eqnarray}
The coefficients for interface F$^b$, $\beta^b_{ij}$, $i=2,3$, are
defined similarly. Notice that $\Gamma^R_2$ and $\tilde{\Gamma}^R_2$
are the ``incoming" amplitudes with respect to F$^b$, so
$\gamma^R_2$ and $\tilde{\gamma}^R_2$ in Eq.
\ref{beta}-\ref{betat} should be replaced by $\Gamma^R_2$ and
$\tilde{\Gamma}^R_2$ in the definition of $\beta^b_{ij}$.
Expressions for $t^a_1$ and $r^a_1$ are obtained from Eq.
\ref{ta2}-\ref{ra2} by index exchange $1\leftrightarrow 2$,
while $t^b_3$ and $r^b_3$ are obtained from Eq.
\ref{tb2}-\ref{rb2} by index exchange $2 \leftrightarrow 3$. The
boundary conditions for the remaining coherence functions, such as
$\tilde{\Gamma}^R_2$ and $\tilde{\gamma}^R_2$, are straightforward
to write down by evoking particle-hole symmetry and are not listed
here \cite{zls}.

\medskip
\begin{figure}[h]
\includegraphics[width=2.5in]{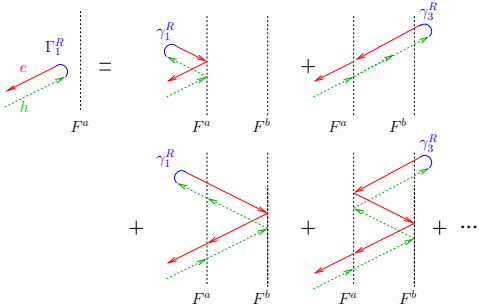}
\caption{Elementary scattering processes contributing to
$\Gamma_1^R$ originating from transmission and reflection from both
interfaces of the FNF trilayer. } \label{diagrams_trilayer}
\end{figure}

The left and right side conducting leads are coupled through the FNF
trilayer. Thus, Ricatti amplitude for one lead depends on elementary
scattering processes that couple excitations between leads $1$ and
$3$. For example, $\Gamma_1^R$ for lead $1$ involves the summation
of multiple scattering processes from both interfaces. Low-order
processes are shown in Fig. \ref{diagrams_trilayer} which include
transmission and reflection from both interfaces. Note that the
processes that couple lead $1$ to $3$ are contained in Eqs.
\ref{GM1}-\ref{rb2}.

Boundary conditions for the distribution functions are also
expressed in terms of the dressed reflection and transmission
amplitudes,
\begin{eqnarray}
X_1^K &= r^a_1\circ x^K_1\circ \overline{r}_1^{a}
    +  t_1^a\circ x^K_2\circ \overline{t}_1^{a}
    -a_1^a\circ \tilde{x}^K_2\circ \overline{a}_1^{a}, \label{X1} \\
X_2^K &= r^a_2\circ x^K_2\circ \overline{r}_2^{a}
    +  t_2^a\circ x^K_1\circ \overline{t}_2^{a}
    -a_2^a\circ \tilde{x}^K_1\circ \overline{a}_2^{a}, \label{X2} \\
x_2^K &= r_2^b\circ X^K_2\circ \overline{r}_2^{b}
    +  t_2^b\circ x^K_3\circ \overline{t}_2^{b}
    -a_2^b\circ \tilde{x}^K_3\circ \overline{a}_2^{b}, \label{x2} \\
X_3^K &= r_3^b\circ x^K_3\circ \overline{r}_3^{b} +  t_3^b\circ
    X^K_2\circ \overline{t}_3^{b} -a_3^b\circ \tilde{X}^K_2\circ
\overline{a}_3^{b}, \label{X3}
\end{eqnarray}
However, the distribution functions also depend on the Andreev
transmission amplitudes, i.e. transmission with branch conversion,
\begin{eqnarray}
a_1^a &= (\Gamma^R_1  \underline{S}^a_{12} - S^a_{12}
    \gamma^R_2)\circ(\tilde{\beta}^a_{22})^{-1},\label{a1a}\\
a_2^a &= (\Gamma^R_2  \underline{S}^a_{21} - S^a_{21}
    \gamma^R_1)\circ(\tilde{\beta}^a_{11})^{-1},\\
a_2^b &= (\gamma^R_2  \underline{S}^b_{23} - S^b_{23}
    \gamma^R_3)\circ(\tilde{\beta}^b_{33})^{-1},\\
a_3^b &= (\Gamma^R_3  \underline{S}^b_{32} - S^b_{32}
    \Gamma^R_2)\circ(\tilde{\beta}^b_{22})^{-1}.\label{a3b}
\end{eqnarray}
They are related to $t$ and $r$ by a set algebraic
identities \cite{zls}. For example, the amplitude $a_1^a$ describing
the event $2h< \rightarrow 1e<$ is related to $r^a_1$ by
$r_1^a=S^a_{11}-a_1^a\tilde\gamma^R_2 S^a_{21}$. Note that both
``retarded'' amplitudes, $r,t$, and the corresponding ``advanced''
counterpart, denoted by $\overline{r},\overline{t}$, enter the
boundary conditions for the distribution functions. These advanced
functions are related to their retarded counterparts by symmetry,
i.e.
$\overline{q}(\epsilon,\epsilon')=q(\epsilon',\epsilon)^{\dagger}$.

\subsection*{Solution to the boundary conditions}

The SFNFS nanopillars are small compared to the superconducting
coherence length. This means that the suppression of the
superconducting order parameter by the exchange fields of the
ferromagnets, as well as the current flow, scales as
$\sqrt{\mathcal{A}}/\xi_0$ and can be neglected. Also the voltage
drop occurs essentially at the contact because of the large Sharvin
resistance. Thus, the values of the ``incoming'' Riccati amplitudes
in the exterior superconducting leads,
$\{\gamma^{R}_j,\tilde{\gamma}^{R}_j,x^K_j,\tilde{x}^K_j; j=1,3\}$,
are determined by the spectrum and distribution functions for bulk
quasiparticle excitations in the left (1) and right (3) leads. Our
task is then to solve the boundary conditions, Eqs.
\ref{GM1}-\ref{a3b}, for the Riccati amplitudes in the N layer
and the ``outgoing'' Riccati amplitudes in the exterior leads.

We start with the coupled Eqs. \ref{GM2}-\ref{gm2}, and solve
them by an iterative method. Typically we are interested in the spin
current in junctions with moderate to high transmission
probabilities, since otherwise the spin-transfer torque effect would
be too small to be of interest. Therefore, a reasonable initial
guess for $\Gamma^R_2$ and $\gamma^R_2$ is obtained by neglecting
normal reflections at F$^a$ and F$^b$. Thus, to zeroth order in
$S^{a/b}_{jj}$, Eqs. \ref{GM2}-\ref{gm2} yield
\begin{eqnarray}
\Gamma^R_2\simeq [(S^{a}_{21})^{\dagger}]^{-1} \gamma^R_1
(\underline{S}^a_{21})^{\dagger}\,, \\
\gamma^R_2\simeq [(S^{b}_{23})^{\dagger}]^{-1} \gamma^R_3
(\underline{S}^b_{23})^{\dagger} \,.
\end{eqnarray}
An improved approximation for $\Gamma^R_2$ and $\gamma^R_2$ is
obtained by substituting the zeroth-order solution into the right
side of Eqs. \ref{GM2}-\ref{gm2}. After the coherence functions
in the N layer are calculated, the outgoing coherence functions in
the leads, $\Gamma^R_1$ and $\Gamma^R_3$, are calculated from Eqs.
\ref{GM1} and \ref{GM3}. The transmission and reflection
amplitudes ($r,t,a$) can then be evaluated from Eqs.
\ref{ta2}-\ref{betat} and \ref{a1a}-\ref{a3b}. Each
iteration of this procedure incorporates higher-order scattering
processes. Operationally the iteration is stopped when the results
for $\Gamma^R_2$ and $\gamma^R_2$ converge within a specified
precision.

The linear equations for the distribution functions, $X^K_2$ and
$x^K_2$, Eqs. \ref{X2}-\ref{x2}, are easily decoupled. For
example, $X^K_2$ obeys the equation
\begin{equation}
X_2^K - c \circ X_2^K \circ \overline{c}=d\,, \label{sylv}
\end{equation}
with $c = r^a_{2} \circ r^b_{2}$ and
\begin{eqnarray}
&d = t^a_{2} \circ {x}^K_1 \circ \overline{t}^a_{2} - a^a_{2} \circ
\tilde{x}^K_1 \circ \overline{a}^a_{2} \nonumber\\
&+ r^a_{2} \circ [ t^b_{2} \circ x^K_3 \circ \overline{t}^b_{2} -
a^b_{2} \circ \tilde{x}^K_3 \circ \overline{a}^b_{2}] \circ
\overline{r}^a_{2}\,.
\end{eqnarray}
Eq. \ref{sylv} is solved iteratively to yield
\begin{equation}
X_2^K=\sum_{n=0}^{\infty}c^n\circ d\circ\overline{c}^n,\;\;c^0=1\,.
\end{equation}
For junctions of moderate to high transparency $c$ is small, and the
series summation converges rapidly. The other distribution functions
in the N layer, $\tilde{X}^K_2$, $x^K_2$, and $\tilde{x}^K_2$, can
be evaluated in a similar way. The distribution functions in the
leads, $X^K_1$ and $X^K_3$, are then calculated from Eq. \ref{X1}
and \ref{X3}.

The main technical difficulty in implementing the iteration
algorithm described above is the evaluation of the folding products.
Following Ref. \onlinecite{arnold} we work in the energy domain. The
folding products are then represented by products of infinite
dimensional matrices for multiple Andreev reflection (MAR). We truncate these
MAR matrices to evaluate the folding products. Details of our analysis
are provided in the appendix.

The observable spin current is calculated from the quasiclassical
Keldysh Green's functions. These functions can be constructed from
the solutions for the Riccati amplitudes \cite{eschrig00},
\begin{eqnarray}\label{gK}
\hat{g}^K &=& \left(
\begin{array}{cc}
g^K & f^K\\
-\tilde{f}^K & -\tilde{g}^K
\end{array}
\right) \\
&=& -(2\pi i)\zeta^R \left(
\begin{array}{cc}
x^K-\gamma^R\tilde x^K\tilde\gamma^A & x^K\gamma^A- \gamma^R\tilde x^K\\
\tilde x^K\tilde\gamma^A -\tilde\gamma^Rx^K & \tilde
x^K-\tilde\gamma^Rx^K\gamma^A
\end{array}
\right) \zeta^A\,, \nonumber
\end{eqnarray}
where
$\zeta^{R,A}=\mathrm{diag}[(1-\gamma^{R,A}\tilde\gamma^{R,A})^{-1},
-(1-\tilde\gamma^{R,A}\gamma^{R,A})^{-1}]$. In particular, the spin
current in the N layer is given by
\begin{equation}
\vec{I} = N_fv_f\mathcal{A}\int_0^{\frac{\pi}{2}}\frac{d\Theta\sin
2\Theta}{4} \int \frac{d\epsilon}{2\pi i}\,
         \mbox{Tr}\left\{
                  \frac{\hbar}{2}\vec{\sigma}\,
                  \left[g^K_{2>}-g^K_{2<}\right]
                  \right\},
\end{equation}
where the trajectories are indicated in Fig. \ref{junction} and
$\Theta=\arccos(\hat{\mathbf{v}}_f\cdot\hat{\mathbf{m}})$ is the
angle between the Fermi velocity and the interface normal
$\hat{\mathbf{m}}$ at F$^a$,
$\vec{\sigma}=(\sigma_x,\sigma_y,\sigma_z)$ are Pauli spin matrices,
and $\mbox{Tr}\{{\ldots}\}$ denotes a trace over spin states.

\subsection*{Interface $\mathbb{S}$ matrix}

The key inputs to our theory are the scattering matrices of the
ferromagnetic layers, $\mathbb{S}^{a}$ and $\mathbb{S}^{b}$. In
principle, these matrices can be calculated from
\textit{ab initio} microscopic theory \cite{xia_ab}.
However, the more efficient
procedure is to fit a few key transport measurements to theoretical
predictions based on models for $\mathbb{S}^{a/b}$. Thus, we proceed
by carrying out calculations with basic models for the
$\mathbb{S}^{a/b}$ matrices which capture the key features of
spin-active scattering at ferromagnetic interfaces, i.e. spin
filtering and spin mixing.
More detailed discussions on the interface $\mathbb{S}$ matrix can
be found in Ref. \cite{zls}.

We parameterize $\mathbb{S}^{a/b}$ in terms of spin-dependent
transparencies, $D_{\uparrow}$ and $D_{\downarrow}$, and a spin
mixing angle, $\vartheta$ \cite{zls}. In particular, for interface
$F^a$,
\begin{equation}
\mathbb{S}^a = \left[\begin{array}{cc} S^a_{11} & S^a_{12} \\
S^a_{21} & S^a_{22} \\ \end{array}\right] = \left[
\begin{array}{cc}e^{-i\phi}S_r &  iS_t\\iS_t &  e^{i\phi}S_r \\ \end{array}
\right]. \label{sa}
\end{equation}
The factors, $e^{\pm i\phi}$, with
$\phi=k_fL/(\hat{\mathbf{p}}_f\cdot\hat{\mathbf{m}})$, describe
the phase of normal-state quasiparticles propagating through the N
layer \cite{note}, while the normal-state spin-dependent reflection
and transmission matrices $S_{r,t}$ are given by
\begin{eqnarray}
S_r&=\left[
\begin{array}{cc}
\sqrt{R}_{\uparrow} e^{i\vartheta/2} & 0 \\
0 & \sqrt{R}_{\downarrow} e^{-i\vartheta/2} \\
\end{array}
\right], \label{sr}\\
S_t&=\left[
\begin{array}{cc}
\sqrt{D}_{\uparrow} e^{i\vartheta/2} & 0 \\
0 & \sqrt{D}_{\downarrow} e^{-i\vartheta/2} \\
\end{array}
\right]. \label{st}
\end{eqnarray}
The reflection and transmission probabilities for the two spin
eigenstates are related by, $R_{\alpha}=1-D_{\alpha}$ for
$\alpha=\uparrow,\downarrow$, defined with respect to the
polarization axis $\hat{\mu}_a$ of interface F$^a$. The
nonequilibrium spin-transfer torque in normal-state magnetic
nanopillars is predominantly determined by the spin filtering effect,
i.e. $D_{\uparrow}\ne
D_{\downarrow}$ \cite{slon,berger,circuit,waintal00,stiles02}. As we
show below, spin mixing becomes important for the nonequilibrium torque
in superconducting magnetic nanopillars. The spin mixing angle,
$\vartheta$, measures the relative phase shift between spin up and
spin down electrons upon transmission or reflection;
$\vartheta$ is easily shown to be the angle of
rotation of the electron spin polarization around the magnetization
direction after the electron is transmitted or reflected. This
process is analogous to Faraday rotation in optics, and in the case
of superconducting leads contributes to the torque on the ferromagnet.
Note that for simplicity we assume the spin mixing angle for reflection
and transmission are the same. This is a feature of models
of the F layer which possess inversion and time reversal 
symmetry (including the inversion of the magnetic moment), see Ref. \cite{zls}.

Since we assumed that F$^b$ and F$^a$ are equivalent except for the
orientation of their moments we can obtain the S-matrix for contact
F$^b$, from $\mathbb{S}^a$ simply by rotating the ferromagnetic
moment by angle $\psi$ with respect to the $x$ axis (see Fig.
\ref{junction}),
\begin{eqnarray}
S^b_{22}&=&e^{i\phi} U(\psi)S_r U(\psi)^{\dagger},\;
S^b_{33}=e^{-2i\phi} S^b_{22},\\
S^b_{23}&=&S^b_{32}=i U(\psi)S_t U(\psi)^{\dagger}, \label{sb}
\end{eqnarray}
where ${U}(\psi)= e^{-i {\sigma}_x\psi/2}$ is the appropriate spin
rotation operator.
We also note that the scattering matrices for holes are also related
to the scattering matrices for electrons by symmetry
relations \cite{millis88},
$\underline{S}^{a/b}_{ij}(\mathbf{p}_{\parallel})=
[S^{a/b}_{ji}(-\mathbf{p}_{\parallel})]^{\mathrm{T}}$, where
the superscript, ${\mathrm{T}}$, denotes the transpose of a spin matrix and
$\mathbf{p}_{\parallel}=\mathbf{p}_{f}-
       (\mathbf{p}_f\cdot\hat{\mathbf{m}})\hat{\mathbf{m}}$.

In general the values of $\{D_{\uparrow},D_{\downarrow},\vartheta\}$
depend on the quasiparticle trajectory ($\hat{\vp}_f$), the specific
ferromagnetic material and the thickness of the F layers. For
example, estimates for the transparencies of the Cobalt layer in
Co-Cu-Co nanopillars are $D_{\uparrow}\sim 0.7$ and
$D_{\downarrow}\sim 0.3$ \cite{waintal00}. We are most interested in
spin-current transfer in relatively high transmission junctions, and
thus relatively weak ferromagnetic materials such as
permalloy \cite{science05,bell}. We consider two representative sets
of spin-dependent transparencies:
\begin{eqnarray}
\mathrm{I.}  & D_{\uparrow}= 0.81, & D_{\downarrow}= 0.64\,, \\
\mathrm{II.} & D_{\uparrow}= 0.95, & D_{\downarrow}= 0.60\,.
\end{eqnarray}
We allow the spin mixing angle to take values ranging between $0$
and $2\pi$, with $\vartheta=\pi$ corresponding to the strongest spin mixing.
This is intended to model F layers of various thickness.
To put these numbers into perspective, consider the delta function
scattering model as an example of weak spin mixing. In this model,
each F layer is approximated by spin-dependent delta function
potentials of strength $V_{\alpha}$ for spin $\alpha$. In terms of
$\theta_{\alpha}\equiv\arctan[\hbar(\mathbf{v}_f\cdot
\hat{\mathbf{m}})/V_{\alpha}]$, the transparencies are
$D_{\alpha}=\sin^2\theta_{\alpha}$, and the spin-mixing angle is
given by $\vartheta=\theta_{\uparrow}-\theta_{\downarrow}=\arcsin
\sqrt{D_{\uparrow}}-\arcsin\sqrt{D_{\downarrow}}$. Thus, for
transparency model I, the delta function barriers yield
$\vartheta=0.061\pi$ (11.0$^{\circ}$); while for model II,
$\vartheta=0.146\pi$ (26.3$^{\circ}$).

In the S matrices defined Eqs. \ref{sa}-\ref{sb}, spin flip
scattering is absent since the ferromagnetic layers are treated as
static macrospins (single domain ferromagnets with homogeneous
magnetization) which give rise only to elastic scattering of
normal-state electrons and holes. Our model is in line with that of
Refs. \onlinecite{waintal00} and \onlinecite{waintal02}. However,
our formalism can be generalized to extend beyond the static
macrospin approximation and to take into account inelastic
scattering and spin flip processes. However, these considerations
are outside the scope of this report.

\subsection*{Transparent spin-mixing}

In the special case where normal reflection and spin filtering at
F$^a$ and F$^b$ are negligible, the boundary conditions are greatly
simplified and can be solved analytically. The nearly transparent
spin-active point contact, introduced in Ref.
\onlinecite{waintal02}, highlights the effect of spin mixing on spin
current transport in magnetic nanopillars. As discussed in Ref.
\onlinecite{zls}, this is a reasonable approximation if the F layers
are made of metallic ferromagnets with exchange energy $h\ll E_f$.
With this approximation, the normal state electron scattering matrix
of F$^a$ is described by only one parameter, the spin-mixing angle,
\begin{equation}
\mathbb{S}^a= \left[
\begin{array}{cc}
  0 & e^{i\hat{\sigma}_{z}\vartheta/2} \\
  e^{i\hat{\sigma}_{z}\vartheta/2} & 0 \\
\end{array}
\right]\,,
\end{equation}
and the scattering matrix for F$^b$ is given by
$\mathbb{S}^b={U}(\psi)\mathbb{S}^a{U}^{\dagger}(\psi)$.

For transparent interfaces, the boundary conditions for Riccati
amplitudes simplify, and we easily obtain the coherence functions in
the N layer,
\begin{eqnarray}
\Gamma^R_2=e^{i\sigma_z\vartheta/2}\gamma^R_1e^{-i\sigma_z\vartheta/2},
\tilde{\Gamma}^R_2=e^{-i\sigma_z\vartheta/2}\tilde{\gamma}^R_1e^{i\sigma_z\vartheta/2}
\,,&&\nonumber \\
\gamma^R_2=e^{i\sigma_{\psi}\vartheta/2}\gamma^R_3e^{-i\sigma'_{\psi}\vartheta/2},
\tilde{\gamma}^R_2=e^{-i\sigma'_{\psi}\vartheta/2}\tilde{\gamma}^R_3e^{i\sigma_{\psi}\vartheta/2}
\,,&& \label{tranGM2}
\end{eqnarray}
where $\sigma_{\psi}=\sigma_z\cos\psi - \sigma_y\sin\psi$ and
$\sigma'_{\psi}=\sigma_{-\psi}$. Similarly we obtain the
distribution functions in the N layer,
\begin{eqnarray}
X^K_2=e^{i\sigma_z\vartheta/2}x^K_1e^{-i\sigma_z\vartheta/2}\,,
\tilde{X}^K_2=e^{-i\sigma_z\vartheta/2}\tilde{x}^K_1e^{i\sigma_z\vartheta/2}
\,,&&\nonumber\\
x^K_2=e^{i\sigma_{\psi}\vartheta/2}x^K_3e^{-i\sigma_{\psi}\vartheta/2}
\,,
\tilde{x}^K_2=e^{-i\sigma'_{\psi}\vartheta/2}\tilde{x}^K_3e^{i\sigma'_{\psi}\vartheta/2}
\,.&&\label{transX2}
\end{eqnarray}

\section{Equilibrium torque}\label{sec_equilibrium-torque}

The equilibrium torque in superconducting nanopillars with chaotic N
spacers has been studied in detail in Ref. \onlinecite{waintal02}.
Here we make explicit connections between the equilibrium torque and
the Andreev bound states in ballistic junctions. The objective is to
reveal the microscopic scattering processes responsible for the
equilibrium spin current, and to facilitate our subsequent
discussion on nonequilibrium torque.

\begin{figure}
\includegraphics[width=3in]{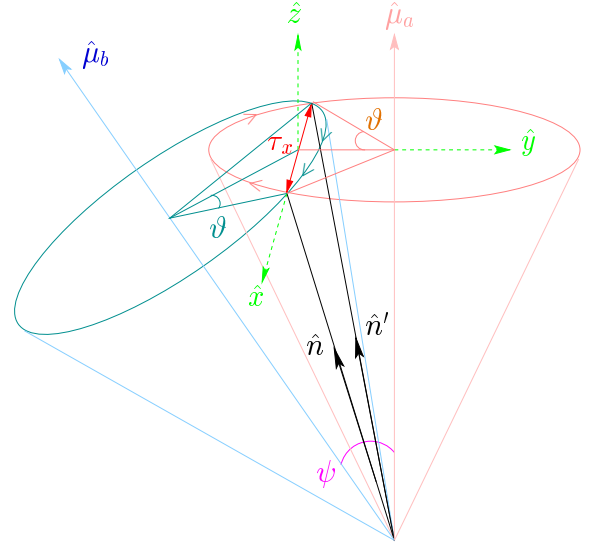}
\caption {The round trip of the spin polarization of a left-moving
electron in the N layer with energy $\epsilon^<_{\pm}$ and initially
polarized along $\hat{\vn}$. Upon traversing F$^a$ to the left
superconductor, $\hat{\vs}$ is rotated from $\hat{\vn}$ by $-\vartheta$
around $\hat{\mu}_a$. After the Andreev reflection at the left
superconductor, the retro-reflected hole passes through F$^a$ again
so that $\hat{\vs}$ is rotated by another $-\vartheta$ around
$\hat{\mu}_a$ and arrives at $\hat{\vn}'$. As a result, spin angular
momentum $\vec{\tau}_a=(\hat{\vn}-\hat{\vn}')\hbar/2$ is
transferred to F$^a$, which as shown is along the $x$ direction. The
right-moving hole passes through F$^b$ and is retro-reflected into
an electron by the right superconductor. After the electron passes
through F$^b$ to get back in N, $\hat{\vs}$ is rotated by
$-2\vartheta$ around $\hat{\mu}_b$ and returns to $\hat{\vn}$. The
second half of the trip transfers spin angular momentum
$\vec{\tau}_b=-\vec{\tau}_a$ to F$^b$.} \label{cone}
\end{figure}

Analytical expressions for the equilibrium spin current for the N
layer of the nearly transparent SFNFS nanopillar are obtained from
the Riccati amplitudes. In particular, the equilibrium Keldysh
Green's functions are simply related to the thermal distribution
function and the retarded Green's functions for the trajectories
2$<$ and 2$>$, i.e. $\hat{g}^R_{2>}$ and $\hat{g}^R_{2<}$, so the
spin current
\begin{equation}
\hspace*{-2mm}
\vec{I}=N_fv_f\mathcal{A}{\hbar\over 2}\int
{d\epsilon\over 2\pi}\tanh\frac{\epsilon}{2T}
\mathrm{Im}\Big\langle\mathrm{Tr}[\vec{\sigma}(g^R_{2>}-g^R_{2<})]\Big\rangle
\,, \label{spin-current_equilibrium}
\end{equation}
where $N_f$ is the density of states at the Fermi energy, $v_f$ is
the Fermi velocity and
$\langle\ldots\rangle\equiv\int^{\pi/2}_{0}\sin\Theta\cos\Theta(\ldots)
d\Theta/2$ is the angular average over the Fermi surface, including
the projection of the quasiparticle spin current along the interface
normal, i.e. $\hat{\vm}\cdot\hat{\vv}_f=\cos\Theta$. The retarded
Green's functions are easily constructed from the Riccati amplitudes
given in Eqs. \ref{tranGM2}. By inspecting the poles of
$g^R_{2<}$, we observe that a pair of Andreev bound states form at
sub-gap energies
\begin{equation}
\epsilon^{<}_\pm(\phi)=
\sgn\left[\sin\left(\frac{\phi\pm\chi}{2}\right)\right]\,
          \cos\left(\frac{\phi\pm\chi}{2}\right)\,\Delta
\,, \label{ABS}
\end{equation}
with
$\chi\equiv\arccos\left[1-2\sin^2\vartheta\cos^2(\psi/2)\right]$.
For trajectory 2$>$, the bound states are at
$\epsilon^{>}_\pm(\phi)=\epsilon^{<}_\pm(-\phi)$. Thus, for a set of
coupled scattering trajectories there are in general four bound
states, $\{\epsilon^B_j$, $j=1,2,3,4\}$. These Andreev bound states
are spin-polarized (c.f. Ref. \onlinecite{bara} for a discussion of
spin-polarized Andreev bound states in SFIFS junctions).
For example, the bound states with energies $\epsilon^{<}_\pm$
correspond to spin eigenstates polarized along the axis
$\hat{\vn}=(n_x,n_y,n_z)$,
\begin{equation}
\hat{\vn}=(\sin{\psi\over2}\sin\vartheta,
          -\sin{\psi\over2}\cos\vartheta,
           \cos{\psi\over2}\cos\vartheta)
\,.
\end{equation}
The Andreev bound states can be visualized as polarized electrons
(holes), confined within the FNF trilayer by the superconducting
pair potential of the two electrodes. Quasiclassically, these bound
states correspond to excitations traversing closed trajectories
between the superconducting leads. Consider an electron on
trajectory 2$<$ for example. In order to form a bound state, its
energy has to be such that the total phase accumulation during a
round trip, including the phase acquired while going through the F
layers as well as that acquired during the Andreev reflection, is a
multiple of $2\pi$ (the Bohr-Sommerfield quantization condition);
additionally its spin has to be polarized along $\hat{\vn}$ so that in
one round trip its spin polarization returns to its original
direction. The evolution of the spin polarization during one round
trip is illustrated in Fig. \ref{cone}.
It follows from the geometry that the mutual torque
between F$^a$ and F$^b$ is along the $x$ direction.

In equilibrium the spin current only flows between the two F layers;
it vanishes in the superconducting electrodes because the total spin
of a Cooper pair is zero (here we only consider conventional
$s$-wave, spin-singlet superconductors). The net contribution from
continuum states to the spin current in Eq.
\ref{spin-current_equilibrium} is zero, so the equilibrium spin
current is carried entirely by the spin-polarized Andreev bound
states. We carry out the integration over the spectrum using the
residue theorem and find that the $y$ and $z$ components of the net
spin current vanish, and that the $x$ component is directly related
to the spectrum of bound states,
\begin{equation}
I_{x}= N_fv_f\mathcal{A}\pi\hbar\sum_{j=1}^4 \Big\langle
{\partial \epsilon^{B}_{j} \over \partial \psi} f(\epsilon^{B}_{j})
\Big\rangle \,, \label{elegant}
\end{equation}
where $f(\epsilon)=1/(e^{\epsilon/k_BT}+1)$ is the Fermi function.
Written explicitly, the spin current in the N layer is given by
\begin{eqnarray}
\vec{I}(\phi,\psi,T)=\hat{x}\,\left(N_fv_f\mathcal{A}\pi\hbar\Delta\right)\,\sin\psi\,
\times\hspace*{3cm}
\nonumber\\
\Big\langle{\sin^2\vartheta\over 2\sin\chi}\sum_{\alpha=\pm
1}\sin{\chi+\alpha\phi\over 2} \tanh\left({\Delta\over
2T}\cos{\chi+\alpha\phi\over 2}\right)\Big\rangle \,.\qquad
\label{fullexp}
\end{eqnarray}
Its magnitude is set by the scale $N_f v_f \mathcal{A}\Delta\hbar$,
where $\Delta$ is the energy gap at temperature $T$. The dependence
of the spin current on $\psi=\arccos(\hat{\mu}_a\cdot\hat{\mu}_b)$
is generally non-sinusoidal since $\chi$ is also a function of
$\psi$. The mutual torque between the F layers,
$\vec{\tau}_b=-\vec{\tau}_a=\vec{I}=\tau \hat{x}$,
is along the $x$ axis, and if the moments were free to rotate, these
torques would drive a mutual precession of $\hat{\mu}_a$ and
$\hat{\mu}_b$.

A similar analysis for the equilibrium Josephson (charge) current,
$I_e$, can be carried out with the result that $I_e$ may be
expressed in terms of the \textsl{phase dispersion} of the Andreev
bound state energies,
\begin{equation}
I_e=2eN_fv_f\mathcal{A}\pi \sum_{j=1}^4
\Big\langle{\partial\epsilon^B_j\over
\partial \phi}f(\epsilon^B_j)\Big\rangle
\,. \label{sup}
\end{equation}
It follows from Eqs. \ref{elegant} and \ref{sup} that
\begin{equation}
\hbar { \partial I_e \over \partial \psi}={2e} {\partial \tau \over
\partial \phi}
\,.
\end{equation}
This relation between the equilibrium spin-transfer torque and the
Josephson charge current was previously written down in Ref.
\onlinecite{waintal00} using free energy arguments. While this
relation provides a nice check on our equilibrium results, it does
not generalize to non-equilibrium charge and spin transport.

Equations \ref{ABS}-\ref{fullexp} are the central results for
the equilibrium spin current. We have explicitly shown that besides
carrying Josephson supercurrent the Andreev bound states mediate a
mutual torque in the $x$ direction between the F layers. As shown in
Fig. \ref{cone}, finite spin mixing at the two F layers, i.e.
rotation of the spin polarization by the exchange field, is crucial
for producing the equilibrium spin-transfer torque. The analysis
outlined above can be carried out for arbitrary SFNFS junctions with
finite spin filtering. The details of the bound state spectrum will be
different, but Eqs. \ref{ABS}-\ref{fullexp} remain valid.

The equilibrium spin-transfer torque in SFNFS junctions and NFNFN
junctions share a similar origin. In both cases, the persistent spin
current flow can be traced to the formation of spin-polarized
bound states and understood as a quantum interference phenomena.
However, there is a major difference. In NFNFN structures, the spin
current is rather fragile to mesoscopic fluctuations
\cite{waintal00}, and strongly suppressed by random impurity
scattering. In reality, the exchange coupling between the F layers
in a normal junction is effectively short-ranged, for example, it
vanishes for $L>20$ nm \cite{tserk}. By contrast, the spin current in
high transmission superconducting junctions is robust against
impurity scattering \cite{waintal02}, and the proximity-induced
exchange coupling is long-ranged. The long range nature of the
induced exchange field in superconductors by a proximity contact
with a ferromagnet was studied in Ref. \onlinecite{toku}.
As demonstrated in normal metal-superconductor hybrid
structures \cite{prox,super}, the phase coherence between incoming
electrons and Andreev reflected holes persists throughout the
diffusive normal metal for energies less than the Thouless energy.
These low energy excitations in the N layer can mediate exchange
coupling between the F layers even when $L$ is comparable to the
superconducting coherence length.

\section{Nonequilibrium torque}\label{sec_nonequilibrium-torque}

An applied voltage across an SFNFS junction leads to time-dependent
non-equilibrium spin currents in the superconducting leads, as well
as the interior normal-metal layer, i.e. currents in all regions of
the trilayer, $\vec{I}^{(i)}(t)$, $i\in\{1,2,3\}$. Since the
Josephson phase is varying as $\varphi(t)=2eVt/\hbar\equiv
\omega_J\,t$ we can expand the currents in each layer in Fourier
series defined by the Josephson frequency,
\begin{eqnarray}
\vec{I}^{(i)}(t) &=& \vec{I}^{(i)}_0(V) \\
                    &+& \sum_{k=1}^{\infty}
               \left( \vec{I}^{(i)}_{kc}(V) \cos (k\omega_J t)
                    + \vec{I}^{(i)}_{ks}(V) \sin(k\omega_J t)\right)\,.
\nonumber
\end{eqnarray}
It follows that the spin-transfer torque on F$^b$,
$\vec{\tau}^b(t)=\vec{I}^{(2)}(t)-\vec{I}^{(3)}(t)$,
has the form
\begin{equation}
\vec{\tau}^b(t)=\vec{\tau}^b_0+\sum_{k=1}^{\infty}
\left(\vec{\tau}^b_{kc} \cos(k\omega_J
t)+\vec{\tau}^b_{ks} \sin(k\omega_J t)\right).
\end{equation}
For the coordinate system shown in Fig. \ref{junction}, the
\textsl{dc} torque $\vec{\tau}^b_0
=\tau^b_{0x}\hat{\vx}+\tau^b_{0y}\hat{\vy}+\tau^b_{0z}\hat{\vz}$ can
be decomposed into the in-plane ($||$) and out of plane ($\perp$)
components, \ber \vec{\tau}_0^{b,||}
&=&\tau^b_{0y}\hat{\vy}+\tau^b_{0z}\hat{\vz}\,,
\\
\vec{\tau}_0^{b,\perp}&=&\tau^b_{0x}\hat{\vx}\,. \eer Note
that both $\vec{\tau}_0^{b,||}$ and
$\vec{\tau}_0^{b,\perp}$ are perpendicular to $\hat{\mu}_b$.
In what follows we discuss how the Fourier components of
$\vec{\tau}_b$ depend on the voltage bias $V$, the spin
mixing angle $\vartheta$, and the misalignment angle $\psi$.

\begin{figure}
  \includegraphics[width=3.4in]{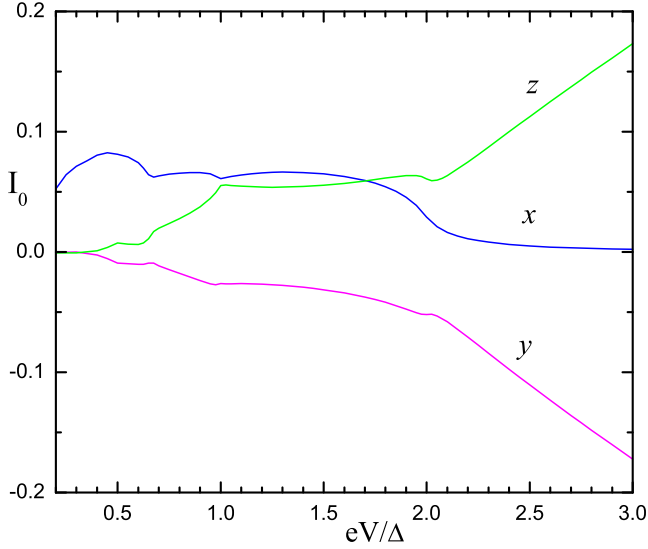}
  \caption{The $x$, $y$, and $z$ component of dc spin current in the N layer for
       perpendicular configuration $\psi=\pi/2$. $D_{\uparrow}=0.81$,
       $D_{\downarrow}=0.64$, $\vartheta=0.061\pi$, $T=0$. The spin mixing is
       rather weak, and the subharmonic gap structures are close to voltage
           $V_n=2\Delta/ne$. The unit for spin current is $N_fv_fA\Delta\hbar/2$.
          }
  \label{dcNxyz}
\end{figure}

\begin{figure}
  \includegraphics[width=3.4in]{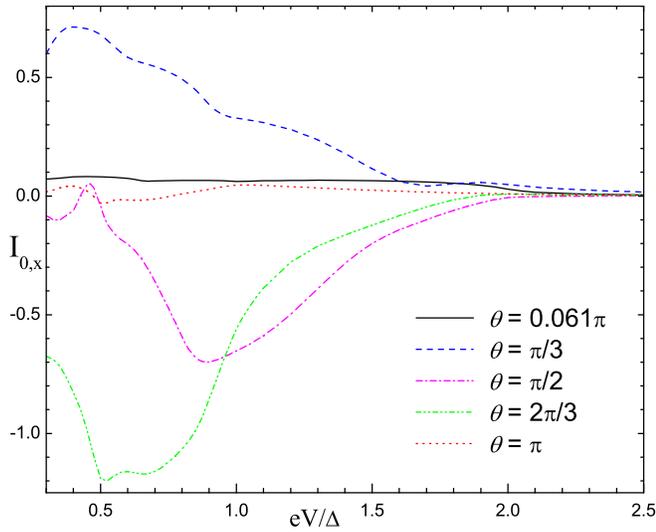}
  \caption{The $x$ component of dc spin current in the N layer for $\psi=\pi/2$
           and different spin mixing angles. $D_{\uparrow}=0.81$,
           $D_{\downarrow}=0.64$, $T=0$.} \label{dcNmix}
\end{figure}

\subsection*{D.C. spin current}

Fig. \ref{dcNxyz} shows the dc spin current for the N layer in the
zero-temperature limit for the delta function scattering model with
$D_{\uparrow}=0.81$, $D_{\downarrow}=0.64$, $\vartheta=0.061\pi$,
and $\psi=\pi/2$. Note that the spin current and spin-transfer
torque are plotted in units of $N_f v_f\mathcal{A}\Delta(T)\hbar/2$
in all figures. We note two major differences between the
nonequilibrium dc spin current in superconducting versus
normal-state magnetic nanopillars. First of all, in normal-state
nanopillars the spin current is linear in the bias
voltage \cite{waintal00} (within the quasiclassical approximation),
whereas for superconducting nanopillars the spin-current-voltage
characteristics are highly nonlinear, and possess subharmonic gap
structure (SGS). Secondly, the spin current in normal-state
nanopillars is predominately due to spin filtering and lies within
the plane spanned by $\hat{\mu}_a$ and $\hat{\mu}_b$ (see Ref.
\onlinecite{waintal00}). By contrast, in superconducting
nanopillars, spin mixing is important; it controls the
current-voltage characteristics of the in-plane spin current, and
leads to a finite out-of-plane spin current along
$\hat{\vx}=\hat{\mu}_a\times\hat{\mu}_b$. As shown in Fig.
\ref{dcNxyz}, the out-of-plane spin current component, $I_{0x}$, is
finite at $V=0$, and decays to zero at high voltages, $eV\gg\Delta$.
On the other hand, the in-plane spin current components, $I_{0y}$ and
$I_{0z}$, are zero at $V=0$ and grow linearly with $V$ at high
voltages.

The SGS of the spin current develops somewhat analogously to the SGS
of dc charge current in voltage-biased Josephson
junctions \cite{arnold,SGS,averin,hamiltonian,zaitsev-averin,andersson}.
The SGS structure reflects the opening of additional channels
(``onsets'') as well as resonances associated with multiple-Andreev
reflection (MAR) processes. The $n$-th order MAR channel has an
onset voltage $V_n=2\Delta(T)/ne$. This transport mechanism is more
efficient for charge transport since the Cooper pairs carry charge
$2e$, but zero spin. A single MAR process may transport multiple
units of charge, but only one quantum of electron spin.
Nevertheless, the sub-gap spin current develops due to MAR, and can
be resonantly enhanced when a MAR trajectory in energy space
intersects one of the FS surface bound states, the energy of which
depends on the spin mixing angle \cite{fogelstrom,andersson}. This
generates resonant transmission of the spin for voltages tuned to
the bound state energy.
%
\begin{figure}[h]
\includegraphics[width=2.5in]{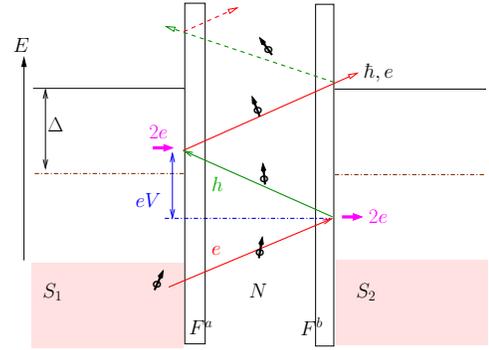}
\caption{Schematic diagram illustrating the trajectory of a sub-gap
excitation
          undergoing multiple Andreev reflection (MAR). Charge transport occurs
          at each Andreev reflection, while spin is transported by the excitation
          with energy above the gap that escapes into the superconducting lead.
         }
\label{MAR_trajectories}
\end{figure}
%
For weak spin mixing, shown in Fig. \ref{dcNxyz}, the SGS exhibits
singularities in the differential conductance ($dI_0/dV$) located
approximately at $V_n$. Increasing the degree of spin mixing,
increases the magnitude of the current, eventually leads to reversal
in the current direction and generally smooths out the
singularities. The spin current vanishes as the spin-mixing angle
approaches $\pi$. Fig. \ref{dcNmix} shows the out-of-plane dc spin
current in the N layer as a function of voltage for a range of spin
mixing angles. For such high transmission junctions, the main
contribution to $I_{0x}$ comes from consecutive spin rotations when
electrons (holes) undergo multiple Andreev reflection. Roughly
speaking, the total number of subgap MARs is inversely proportional
to $V$; as a result $I_{0x}$ decays rapidly at high voltages. The
magnitude of $I_{0x}$ also becomes small for $\vartheta$ close to
$0$ or $\pi$, where the quasiparticle spin direction is hardly
changed in each Andreev reflection (see Fig. \ref{cone}).

\subsection*{D.C. spin-transfer torque}

\begin{figure}
  \includegraphics[width=3.4in]{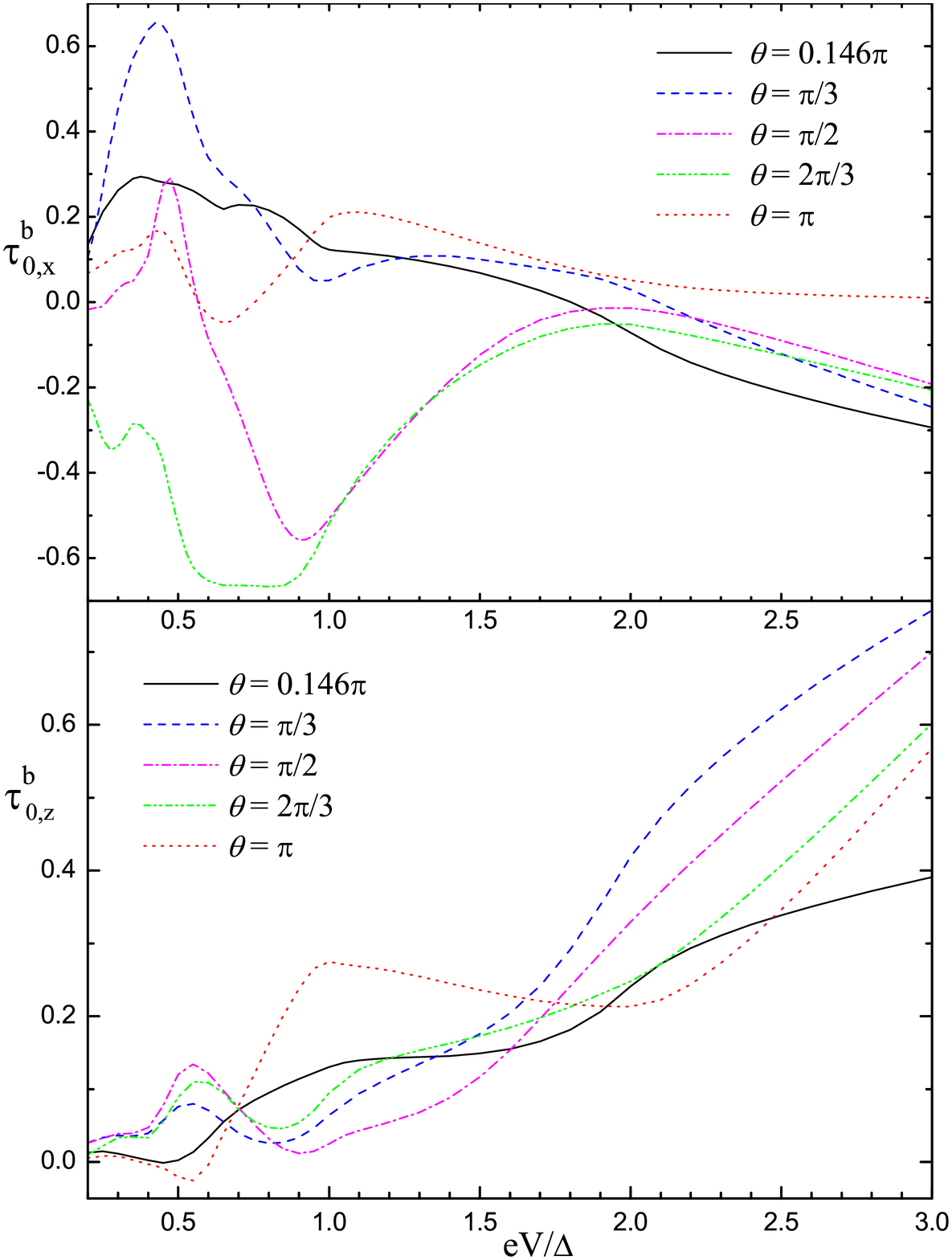}
  \caption{The dc torque on F$^b$ in $x$ (upper panel) and $z$ (lower panel)
  direction for $\psi=\pi/2$ and different spin mixing angles.
  $D_{\uparrow}=0.95$, $D_{\downarrow}=0.6$, $T=0.5T_c$.}
  \label{compare}
\end{figure}

Once the spin current is known in each region, the torque on F$^b$
is obtained by
$\vec{\tau}^b=\vec{I}^{(2)}-\vec{I}^{(3)}$. Fig.
\ref{compare} shows the dc torque on F$^b$ as a function of the bias
voltage for $D_{\uparrow}=0.95$, $D_{\downarrow}=0.6$, $\psi=\pi/2$,
$T=0.5T_c$, and a variety of spin mixing angles. Note that the $y$
component of $\vec{\tau}^b$ vanishes for $\psi=\pi/2$. We
observe that the voltage dependence of the in-plane torque
($\tau^b_{0z}$) and the out-of-plane torque ($\tau^b_{0x}$) are
sensitive to spin mixing, and more importantly, exhibit very
different voltage characteristics. The in-plane torque vanishes for
$V\rightarrow 0$, while the out-of-plane torque is finite at zero
voltage.
The out-of-plane torque varies more dramatically with the bias
voltage for $V<2\Delta/e$, changing sign around $2\Delta$ in the
case of weak spin mixing (i.e. $\vartheta<\pi/3$). This marks the
crossover from the low-voltage regime, where only high-order MAR
processes with finite number of subgap Andreev reflections
contribute to the spin momentum transfer, to the high-voltage
regime, where direct transmission without Andreev reflection
dominates the spin transport. Indeed, for $V\gg 2\Delta$, both
components become linear in $V$, however, note that the magnitude of
$\tau^b_{0z}$ (for $V\gg 2\Delta$) is suppressed from its
normal-state value because of the superconducting gap in the
excitation spectrum. Finally, we note that that if we scale the
torque in units of $N_f v_f \mathcal{A}\Delta(T)\hbar/2$ and the
voltage in units of $\Delta(T)/e$ then both in- and out-of-plane
components of the torque are nearly temperature
independent.

\begin{figure}
  \includegraphics[width=3.4in]{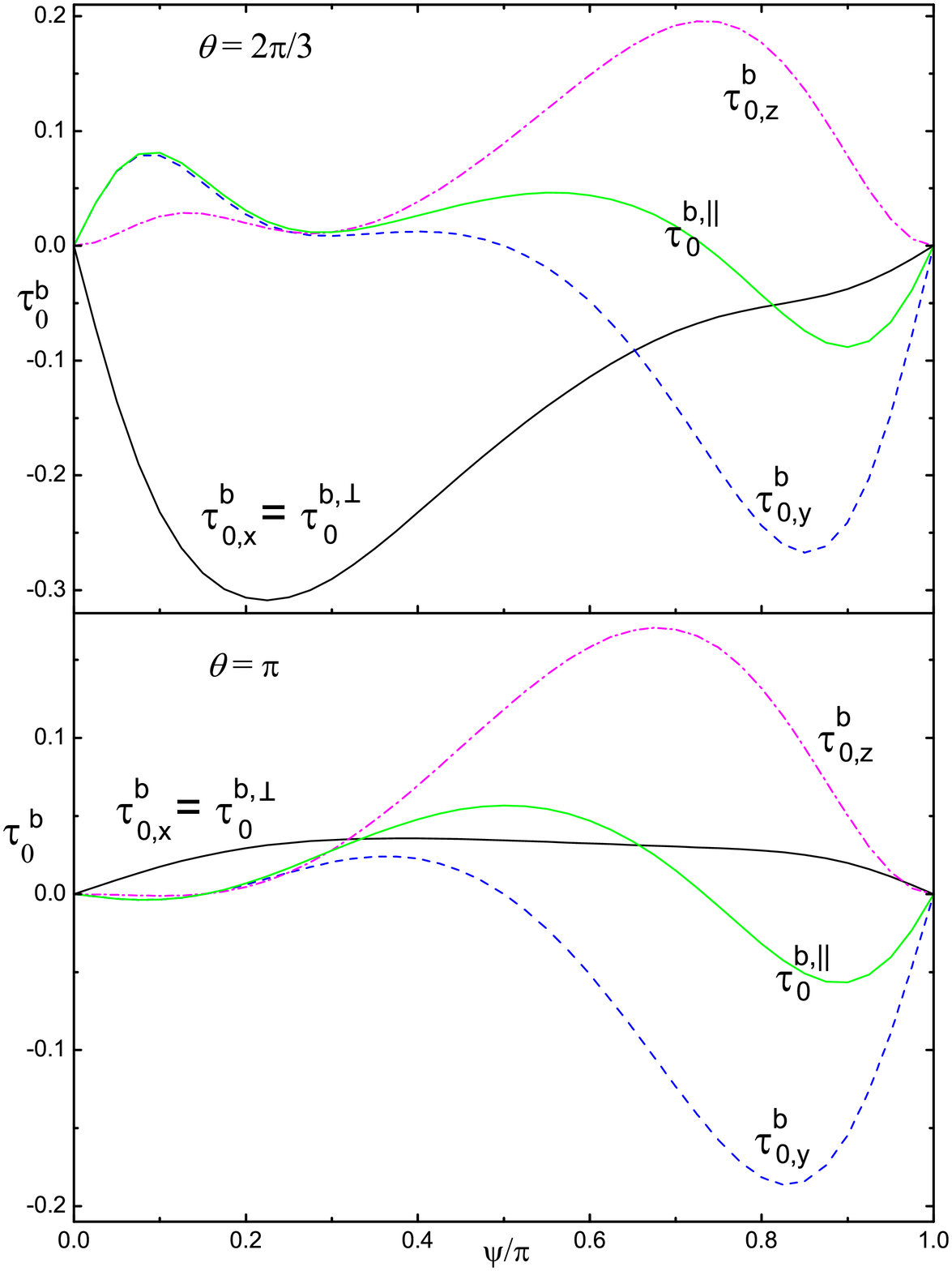}
  \caption{The dc torque on F$^b$ as a function of misalignment angle
  $\psi$ for strong spin mixing, $\vartheta=2\pi/3$ and
  $\pi$. $D_{\uparrow}=0.81$, $D_{\downarrow}=0.64$, $T=0.5T_c$, $V=1.5\Delta/e$.
  The $x$, $y$, and $z$ component of the dc torque are plotted
  separately. Also shown is the magnitude of the in-plane torque (labeled by
  ``in"), the vector sum of the $y$ and $z$ components.}
  \label{psi}
\end{figure}

\begin{figure}
  \includegraphics[width=3.4in]{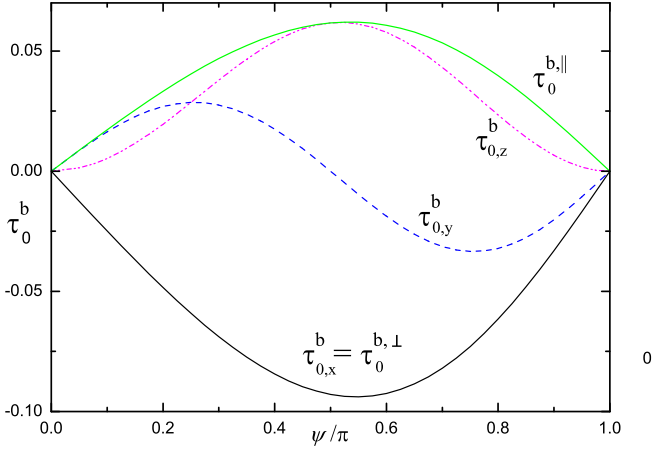}
  \caption{
  The dc torque on F$^b$ as a function of misalignment angle $\psi$ for weak spin mixing,
  $\vartheta=0.061\pi$, at high voltages, $V=2.5\Delta/e$.
  $D_{\uparrow}=0.81$, $D_{\downarrow}=0.64$, $T=0.5T_c$.
  } \label{angle}
\end{figure}

The spin-transfer toque vanishes in the parallel and antiparallel
configurations for the two F layers, i.e. $\psi=0$ and $\pi$. The
functional forms for $\vec{\tau}^{b,||}_{0}(\psi)$ and
$\vec{\tau}^{b,\perp}_{0}(\psi)$ are generally complicated.
Two calculations are shown in Fig. \ref{psi} for $\vartheta=2\pi/3$
and $\vartheta=\pi$ at voltage $V=1.5\Delta/e$, with
$D_{\uparrow}=0.81$, $D_{\downarrow}=0.64$, and $T=0.5T_c$. In the
case of $\vartheta=2\pi/3$, the magnitude of out-of-plane torque is
considerably larger than the in-plane torque, and it possesses a
pronounced maximum around $\psi\simeq \pi/5$. In the case of
$\vartheta=\pi$, however, the out-of-plane torque is of the same
order as the in-plane torque, and it varies rather slowly with
$\psi$. At higher voltage, $eV > 2\Delta$, and weak spin mixing both
the in- and out-of-plane torque vary approximately as $\sin\psi$, as
shown in Fig. \ref{angle} for $\vartheta=0.061\pi$ and
$V=2.5\Delta/e$. The high-voltage asymptotic behavior of
$\vec{\tau}^{b,||}_{0}(\psi)$ is similar to that of the
normal-state magnetic nanopillars discussed in Ref.
\onlinecite{waintal00}.

\subsection*{A.C. spin-transfer torque}

\begin{figure}
  \includegraphics[width=3.4in]{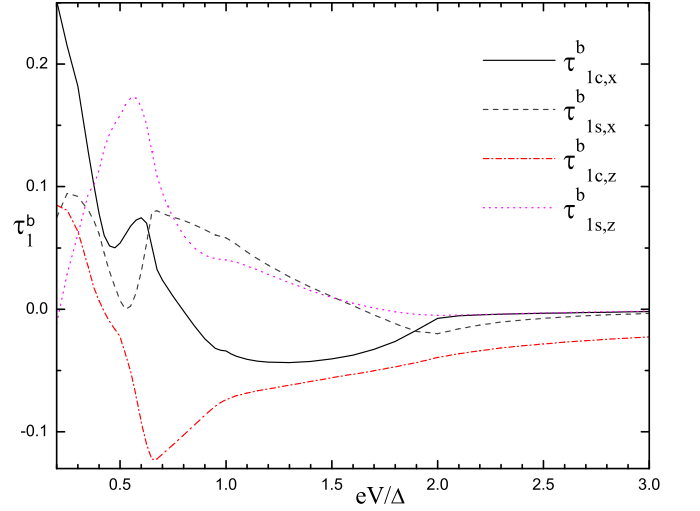}
  \caption{ The first Fourier component of the ac torque on F$^b$.
  Shown are $\vec{\tau}_{1c}$ and $\vec{\tau}_{1s}$ in $x$ and $z$ direction
  (the $y$ components are zero for $\psi=\pi/2$).
  $D_{\uparrow}=0.95$, $D_{\downarrow}=0.6$, $\vartheta=0.146\pi$ (weak spin mixing), $\psi=\pi/2$, and $T=0.5T_c$.
  } \label{tor}
\end{figure}

Figure \ref{tor} shows the first Fourier components of the ac torque
on F$^b$, $\vec{\tau}_{1c}(V)$ and
$\vec{\tau}_{1s}(V)$, for $D_{\uparrow}=0.95$,
$D_{\downarrow}=0.6$, $\vartheta=0.146\pi$, $\psi=\pi/2$, and
$T=0.5T_c$. The ac spin current comes from the interference between
MAR processes of different order. As a result, the ac torque rapidly
decays to zero at high voltages. To get a better understanding of
the ac torque, in what follows we derive its analytical expression
for transparent SFNFS point contacts.

\begin{figure}
\includegraphics[width=3.4in]{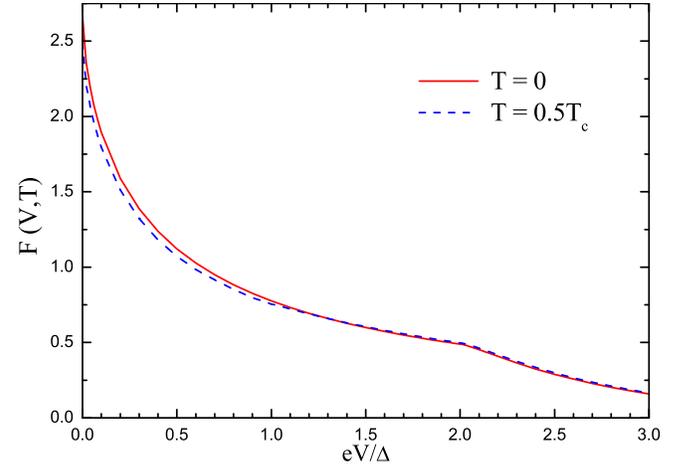}
\caption {Dimensionless factor, $F(V,T)$, for the torque in Eq.
\ref{volt} for $T=0$ and $T=0.5T_c$. $F(V,T)$ is
inversely proportional to $V$ for $eV\gg\Delta$.} \label{tran-sin}
\end{figure}

The nonequilibrium spin current in transparent point contacts can be
computed directly from Eqs. \ref{tranGM2}-\ref{transX2}. The
analysis is carried out in the appendix; the resulting expression
for $\vec{I}^{(2)}$ is given by Eq. \ref{trannon}. We observe
that the dc spin current vanishes in all three regions, $i=1,2,3$.
This is a feature of the nearly transparent interface model, and
it shows that finite spin filtering is required for a non-zero dc
torque. For the ac current the dominant contribution comes from the
first Fourier component, $\vec{I}^{(i)}_{1c}(V)\cos(\omega_J
t)+\vec{I}^{(i)}_{1s}(V)\sin(\omega_J t)$. The cosine term is
negligible compared to the sine term when the inelastic rate is
small; we have assumed $\Gamma_{\mathrm{in}}=10^{-4}\Delta$.
The sine component of the current vanishes in the superconducting
leads, however $\vec{I}^{(2)}_{1s}$ is finite and polarized along
the $\hat{\vx}$. Thus, the time-dependent torque is
\begin{equation}
\vec{\tau}^b(t)=-\vec{\tau}^a(t)\simeq
                        \tau_{1s}(V,T)\sin(\omega_J t)\hat{\vx}
\,.
\end{equation}
From Eq. \ref{trannon}, we find the torque is proportional to
$\sin^2\vartheta$ and $\sin\psi$,
\begin{equation}
\tau_{1s}(V,T)=N_f v_f\mathcal{A}\Delta(T){\hbar\over 2}\, F(V,T)
\Big\langle\sin^2\vartheta\Big\rangle\,\sin\psi \,. \label{volt}
\end{equation}
The voltage dependence of $\vec{\tau}^b$ is given by the
dimensionless function $F(V,T)$, plotted in Fig. \ref{tran-sin}
for $T=0.5T_c$ and $T=0$. Since $F(V,T)$ depends weakly on $T$,
the temperature dependence of $\vec{\tau}_b$ is mainly
determined by the gap $\Delta(T)$.

Note that the ac torque decays monotonically with voltage in
transparent contacts. This is in sharp contrast with the ac torque,
$\tau^b_{1s}(V)$, shown in Fig. \ref{tor} for $D_{\alpha}<1$. In
this case resonant MAR leads to sharp structure in
$\tau^b_{1s}\hat{\vx}$ for $V<2\Delta$. At high voltage,
$eV$$\gg$$\Delta$, the dominant contribution to spin current comes
from the lowest order MAR processes, i.e. the terms $m=0,1$ in Eq.
\ref{trannon}. Keeping only the leading order terms we obtain the
asymptotic behavior for $F(V,T)$ for $D_{\alpha}=1$ given by
\begin{eqnarray}
F(V,T)\simeq 2\mathrm{Re}\int_{-\infty}^{+\infty}
{d\epsilon\over\Delta} \Big\langle
x^K_3(\epsilon_1)\gamma^R_3(\epsilon_{-1})\tilde{\gamma}^R_1(\epsilon)
\nonumber\\
-x^K_1(\epsilon)\gamma^R_1(\epsilon_{2})\tilde{\gamma}^R_3(\epsilon_{1})
\Big\rangle \,. \label{ac-current_amplitude}
\end{eqnarray}
At zero temperature integration of Eq. \ref{ac-current_amplitude} yields
\begin{equation}
\tau_{1s}(T=0)\simeq N_f v_f\mathcal{A}\Delta\hbar{\Delta\ln 2\over
eV} \Big\langle\sin^2\vartheta\Big\rangle\,\sin\psi \,,
\end{equation}
and thus, the ac spin-transfer torque is inversely proportional to
the voltage bias in the limit $eV$$\gg$$\Delta$.

\begin{figure}
\includegraphics[width=3.4in]{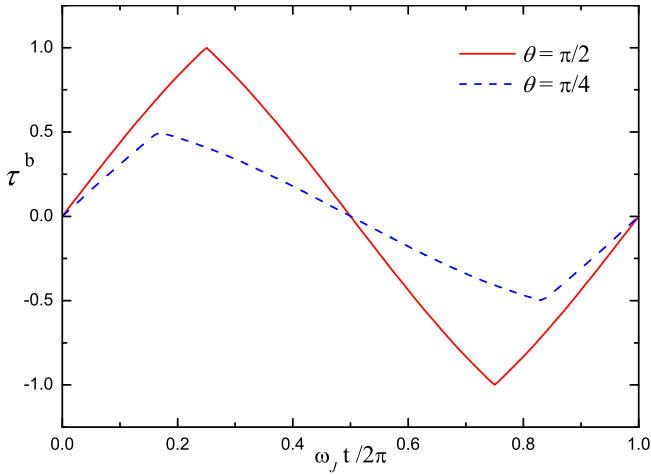}
\caption{The time evolution of the spin-transfer torque on F$^b$ in
the adiabatic limit. $\psi=\pi/2$, $T=0$. Although
$\tau_{1s}\sin(\omega_Jt)$ is the dominant term in the Fourier
expansion of $\tau^b(t)$, the deviation from the $\sin(\omega_Jt)$
dependence is obvious. } \label{adiabatic}
\end{figure}

In the opposite limit, $eV$$\ll$$\Delta$, the time evolution of
spin-transfer torque is governed by the nearly adiabatic dynamics of
the Andreev bound states. The spin-transfer torque, however, does
not assume its instantaneous equilibrium value, i.e.
$\vec{\tau}(t)\neq \vec{\tau}_{\mathrm{eq}}(\phi(t))$.
The reason is that the \textsl{dynamics} of the bound state
\textsl{spectrum}, i.e. $\epsilon^B_j(\phi(t))$, leads to out
of equilibrium population of the Andreev bound states. The
occupation of the Andreev bound states remains constant as a
function of time until the bound state energy evolves to the gap
edge at which point these states equilibrate with the continuum
quasiparticles in the leads within a time scale much shorter than
the period of a Josephson oscillation. This argument was made by
Averin et al. to describe the adiabatic time evolution of the charge
transport in high transmission Josephson point
contacts \cite{averin,abpb}. To obtain the adiabatic limit for the
dynamics of the spin-transfer torque, we replace the Fermi function
$f(\epsilon_j)$ in Eq. \ref{elegant} with $f(\pm\Delta)$. For
$eV>0$ the Josephson phase $\varphi$ increases with time. The phase
dispersion of Andreev bound states given by Eq. \ref{ABS}
indicates that the occupation of bound states at energy
$\epsilon^<_{\pm}(\phi(t))$ is given by $f(\Delta)$ while the
occupation of bound states at energy $\epsilon^>_{\pm}(\phi(t))$ is
given by $f(-\Delta)$. This procedure yields
\begin{eqnarray}
\vec{\tau}^b(t)\simeq N_f v_f\mathcal{A}\Delta\pi{\hbar\over
2}\tanh{\Delta\over 2T} \hspace*{3cm}
\nonumber\\
\Big\langle {\sin^2\vartheta\over\sin\chi} \left[|\sin({\chi\over
2}+\omega t)|-|\sin({\chi\over 2}-\omega t)|\right] \Big\rangle
\hat{\vx}\,.
\end{eqnarray}
The time evolution of the torque, $\tau^b(t)$, at zero temperature
is shown in Fig. \ref{adiabatic} for $\psi=\pi/2$, $\vartheta=\pi/2$
and $\pi/4$. If we expand $\tau_b(t)$ in Fourier series, we find the
sine part of the first Fourier component is
$\tau_{1s}=\frac{4}{3}\,N_fv_f\mathcal{A}\Delta\hbar$ for
$\psi=\vartheta=\pi/2$, which agrees with our numerical result in
Fig. \ref{tran-sin}. It is clear that the higher order Fourier
components are nonzero, but their magnitudes are much smaller than
$\tau_{1s}$.

\section{Summary}\label{sec_discussion}

In this paper we studied spin momentum transfer in
superconducting nanopillars in the clean limit using models for the
interface scattering matrices for the magnetic contacts. Our method presented in section
\ref{sec_methods}, however, is quite general. Further exploration of
the parameter space $\{D_\alpha,\vartheta,\psi\}$ as well as
investigations on the effects of impurity scattering in the N
spacer, spin-flip scattering at the ferromagnetic interfaces, etc.,
can be carried out within the current theoretical framework. The
available information on the scattering parameters for SFNFS
structures, e.g. the directional dependence of spin mixing angles,
is a potential limitation. However, spin transport is dominated by
trajectories close to normal incidence.
For trajectories substantially away from the normal direction, e.g.
the grazing trajectories $\Theta\sim\pi/2$, the effective thickness
of the F layer is large and the transmission probability is very small.
Their contribution to the total transport current is negligible
even though the spin mixing might be strong.
Therefore we believe that the basic features we
obtain from our S-matrix models for the F interfaces, such as the
out-of-plane torque, its nonlinear voltage dependence, etc. will
survive the average over scattering trajectories for more detailed
models for the interface S matrices. Indeed we expect our model to
provide a good approximation for realistic device behavior.

In summary, we investigated the nonequilibrium spin-transfer torque
in voltage-biased superconducting magnetic nanopillars. Our work
extends earlier research on equilibrium phase-sensitive
spin-transfer torque by Waintal and Brouwer \cite{waintal02} to
nonequilibrium junctions. Our results, and the general theoretical
framework, provide a computational formalism for understanding the
operation of non-equilibrium superconducting spin-transfer devices.
We have shown that a superconducting nanopillar is an interesting
system with new physics, particularly a rich dynamics resulting from
the interplay of multiple Andreev reflection, spin mixing, spin
filtering, spectral dynamics of the interface states and the
Josephson phase dynamics. The nonequilibrium torque at finite
voltages is in principle observable by monitoring the magnetization
dynamics, as described in Ref. \onlinecite{waintal02}. However, the
details of the coupled magnetization dynamics and the time-dependent
spin-transfer torque is a topic for a future study.

\begin{acknowledgements}
We thank Dr. Tomas L\"ofwander for stimulating discussions on this work,
and for the diagram notation of the interface boundary conditions used in
Sec. \ref{sec_methods}.
\end{acknowledgements}

\appendix
\section{The folding product}

Our evaluation of folding products follows closely the pioneering
work of Arnold \cite{arnold}. Double time correlation functions can
be Fourier transformed between the time domain and the energy
domain,
\be A(t_1,t_2)=\int_{-\infty}^{+\infty}{d\epsilon'\over
2\pi}\int_{-\infty}^{+\infty}{d\epsilon''\over
2\pi}e^{-i\epsilon't_1}A(\epsilon',\epsilon'')e^{i\epsilon''t_2} \,.
\ee
The folding product is most conveniently evaluated in the energy
domain,
\be [A\circ
B](\epsilon',\epsilon'')=\int_{-\infty}^{+\infty} {d\epsilon\over
2\pi}A(\epsilon',\epsilon) B(\epsilon,\epsilon'') \,.
\ee
We can fix
the condensate phase and electrical potential of the left electrode
to be zero. It is easy to show that the matrix elements of
``incoming" Riccati amplitudes in the electrodes are given by
\begin{eqnarray}
\gamma^R_1(\epsilon',\epsilon)&=&2\pi\delta(\epsilon'-\epsilon)\gamma^R_1(\epsilon)\,,\\
x^K_1(\epsilon',\epsilon)&=&2\pi\delta(\epsilon'-\epsilon)x^K_1(\epsilon)\,,\\
\gamma^R_3(\epsilon',\epsilon)&=&2\pi\delta(\epsilon'-\epsilon+2\omega)\gamma^R_3(\epsilon-\omega)\,,\\
x^K_3(\epsilon',\epsilon)&=&2\pi\delta(\epsilon'-\epsilon)x^K_3(\epsilon+\omega)
\,.
\end{eqnarray}
where $\omega=\omega_J/2$, and $\gamma^R(\epsilon)$ etc. take their
bulk equilibrium value \cite{eschrig00}. The remaining Riccati
amplitudes are obtained from those above by the symmetry relation,
$\tilde{q}(\hat{p}_f,\epsilon',\epsilon)\equiv
q^*(-\hat{p}_f,-\epsilon',-\epsilon)$. Note that $\gamma^R_3$ and
$\tilde{\gamma}^R_3$ are ladder operators in energy space, while the
amplitudes $\gamma^R_1, \tilde{\gamma}^R_1, x^K_j$ and
$\tilde{x}^K_j$ are diagonal in energy space.

Inspection of the boundary conditions, Eqs. \ref{GM1}-\ref{a3b},
shows that all the coherence functions in the N layer, as well as
all the effective scattering amplitudes ($r,t,a$) defined at F$^a$
and F$^b$, have the form of \be
A(\epsilon',\epsilon)=\sum_{m=-\infty}^{+\infty}2\pi
\delta(\epsilon'-\epsilon_m)A_{m0}(\epsilon) \,, \ee where we
introduced the short-hand notation,
$\epsilon_m\equiv\epsilon+m\omega$. This expansion reflects the fact
that quasiparticles of energy $\epsilon$ are only scattered to
side-band energies, $\{\epsilon_m\}$, during multiple Andreev
reflection processes. Furthermore, the matrix elements of the
distribution functions in the N layer are nonzero only between
states corresponding to the side band energies,
\be
X(\epsilon',\epsilon'')=\sum_{m,k}\int {d\epsilon\over 2\pi}
2\pi\delta(\epsilon'-\epsilon_{2m})
2\pi\delta(\epsilon''-\epsilon_{2k})X_{2m,2k}(\epsilon) \,.
\ee
The
integration variable $\epsilon$ provides a natural reference energy
to introduce a discrete basis, $\{\left|m\right>\equiv
\left|\epsilon_m\right>\}$, in energy space \cite{arnold},
\be \sum_m
\ket{m}\bra{m}=1,\;\; A_{mn}(\epsilon)\equiv \bra{m}A\ket{n} \,.
\ee
In the Arnold basis, each spectral operator is represented by an
infinite dimensional matrix, and the folding product reduces to
matrix product, \be [A\circ
B]_{mn}(\epsilon)=\sum_{k}A_{mk}(\epsilon) B_{kn}(\epsilon) \,. \ee
For general interface scattering matrices, the iteration algorithm
described in Sec. \ref{sec_methods} to solve the boundary conditions
must be carried out numerically. To this end, we truncate the
infinite dimensional matrices in energy space and work in a space of
dimension (2M+1)$\times$(2M+1). This corresponds to neglecting
contributions from MAR processes with more than M
reflections \cite{arnold}. Usually M$>$4 yields sufficiently accurate
results except for small voltages, e.g. $eV<0.1\Delta$.

Once the Riccati amplitudes are calculated using the iteration
algorithm, we can assemble the Keldysh Green's functions to obtain
the spin current, e.g. in the N layer,
\begin{eqnarray}
\vec{I}^{(2)}(t,V)= N_f v_f\mathcal{A}\pi{\hbar\over 2}\,
\Re\sum_{k=-M}^{M}\sum_{m}^{|k+m|\leq M}e^{i2m\omega t}
\nonumber\\
\times \int {d\epsilon\over 2\pi}
\Big\langle\mathrm{Tr}\left[\vec{\sigma}\mathrm{K}_{(k,k+m)}(\epsilon)\right]\Big\rangle
\,,
\end{eqnarray}
where the kernel is given \be \mathrm{K}_{(m,n)}(\epsilon)\equiv
{1\over 2\pi i} \bra{2m}g^K_{2>}-g^K_{2<}\ket{2n} \,. \ee The spin
currents in the left and right electrodes, $\vec{I}^{(1)}(t)$ and
$\vec{I}^{(3)}(t)$, are computed in similar fashions.

For transparent SFNFS point contacts, the spin current can be
obtained analytically. In constructing the Keldysh Green's
functions, the matrix elements of $(1-A)^{-1}$ are computed by
expanding it in geometric series, $
(1-A)^{-1}=\sum_{k=0}^{\infty}A^k$. For example, the probability
amplitude of bulk quasiparticles being scattered from energy
$\epsilon$ to $\epsilon+2k\omega$ via MAR is
\begin{eqnarray}
\mathrm{M}_{k}(\epsilon)=
\left<2k|(1-\Gamma^R_2\odot\tilde{\gamma}^R_2)^{-1}|0\right>
=
\hspace*{2.5cm}&&
\\
\prod_{j=1}^{k}e^{i\sigma_z\vartheta/2}\gamma^R_1(\epsilon_{2j})e^{-i\sigma_z\vartheta/2}
e^{-i\sigma'_{\psi}\vartheta/2}\tilde{\gamma}^R_3(\epsilon_{2j-1})e^{i\sigma_{\psi}\vartheta/2}.
&&
\nonumber
\end{eqnarray}
We adopt the convention $\prod_{j=1}^{0}(...)=1$ and
$\prod_{j=1}^{k>j}A_j=A_kA_{k-1}...A_2A_1$. Besides M$_k$, the
following transition amplitudes are required to construct
$\hat{g}^K_{2>}$ and $\hat{g}^K_{2<}$,
\begin{eqnarray}
\mathrm{N}_{k}(\epsilon)&=&
\mathrm{M}_{k}(\epsilon)e^{i\sigma_z\vartheta/2}\gamma^R_1(\epsilon)e^{-i\sigma_z\vartheta/2},
\\
\mathrm{O}_{k}(\epsilon)&=&
\prod_{j=1}^{k}e^{i\sigma_{\psi}\vartheta/2}
\gamma^R_3(\epsilon_{-2j+1}) e^{-i\sigma'_{\psi}\vartheta/2}
e^{-i\sigma_z\vartheta/2}
\nonumber\\
&& \times \tilde{\gamma}^R_1(\epsilon_{-2j+2})
e^{i\sigma_z\vartheta/2},\\
\mathrm{P}_{k}(\epsilon)&=&
\mathrm{O}_{k}(\epsilon)e^{i\sigma_{\psi}\vartheta/2}\gamma^R_3(\epsilon_1)e^{-i\sigma_{-\psi}\vartheta/2}
\,.
\end{eqnarray}
The magnitudes of these amplitudes decay rapidly with increasing MAR
order, i.e. $k$, as well as the bias voltage. The spin current in
the N layer is found to be
\begin{eqnarray}
\vec{I}^{(2)}(t,V)&=& N_f v_f\mathcal{A}\pi{\hbar\over 2}\,
\Re\sum_{m=-\infty}^{+\infty}\sum_{k=0}^{\infty}e^{i2m\omega t}
\nonumber \\ &&\times \int {d\epsilon\over 2\pi}
\left<\mathrm{Sp}[\vec{\sigma} \mathrm{K}_{(k,m)}(\epsilon)]\right>,
\label{trannon}
\\
\mathrm{K}_{(k,m)}&=&\mathrm{N}_k(\epsilon)
\tilde{x}^K_3(\epsilon_{-1})\mathrm{N}^{\dagger}_{k+m}(\epsilon)
-\mathrm{M}_k(\epsilon)
x^K_1(\epsilon)\mathrm{M}^{\dagger}_{k+m}(\epsilon)
\nonumber\\
&+&\mathrm{O}_k(\epsilon)
x^K_3(\epsilon_1)\mathrm{O}^{\dagger}_{k-m}(\epsilon)-\mathrm{P}_k(\epsilon)
\tilde{x}^K_1(\epsilon_2)\mathrm{P}^{\dagger}_{k-m}(\epsilon) \,.
\nonumber
\end{eqnarray}
$\vec{I}^{(1)}$ and $\vec{I}^{(3)}$ have the same form as Eq.
\ref{trannon}, but with slightly different definitions of the
transition amplitudes
$\{\mathrm{M}_k,\mathrm{N}_k,\mathrm{O}_k,\mathrm{P}_k\}$.

\end{document}